\documentclass[conference]{IEEEtran}

\ifCLASSINFOpdf
\else
\fi
\newcommand{\thesystem}{HotFuzz\xspace}
\newcommand{\fuzzer}{$\mu$Fuzz\xspace}
\newcommand{\eyevm}{Eye\-VM\xspace}
\newcommand{\ivi}{IVI\xspace}
\newcommand{\sri}{SRI\xspace}

\newcommand{\zerodaynum}{26\xspace}
\newcommand{\mavenzerodaynum}{132\xspace}
\newcommand{\vulnmavenlibraries}{47\xspace}
\newcommand{\jninetime}{4 and a half months\xspace}
\newcommand{\javacount}{Fifteen billion\xspace}

\newcommand{\methodundertest}{method under test\xspace}

\newcommand{\jreversion}{1.8.0\_181\xspace}
\newcommand{\jreemptybugs}{6\xspace}

\newcommand{\rdtsc}{\texttt{rdtsc}\xspace}
\newcommand{\hotfuzzsloc}{5,487\xspace}
\newcommand{\eyevmsloc}{1,007\xspace}
\newcommand{\synthsloc}{288\xspace}
\newcommand{\tojsonobjecttime}{120 hours\xspace}
\newcommand{\totalbugs}{158\xspace}

\newcommand{\stacchallenges}{80\xspace}
\newcommand{\stac}{DARPA Space and Time Analysis for Cybersecurity (STAC)\xspace}
\newcommand{\resplittime}{7 days\xspace}
\newcommand{\solvedchallenges}{5\xspace}
\newcommand{\totalchallenges}{36\xspace}
\newcommand{\jremethodscovered}{29.17\%\xspace}
\newcommand{\stacmethodscovered}{32.96\%\xspace}
\newcommand{\mavenmethodscovered}{29.66\%\xspace}
\newcommand{\library}[1]{\textit{#1}\xspace}
\newcommand\scalemath[2]{\scalebox{#1}{\mbox{\ensuremath{\displaystyle #2}}}}

\usepackage[hidelinks]{hyperref}
\usepackage{amsmath}
\usepackage{epsfig}
\usepackage{float}
\usepackage{endnotes}
\usepackage{paralist}
\usepackage{shortcuts}
\usepackage{nicefrac}
\usepackage{multirow}
\usepackage{tabularx}
\usepackage{tabulary}
\usepackage{mdwlist}
\usepackage{listings}
\usepackage{booktabs}
\usepackage{balance}
\usepackage{multirow}
\usepackage{parskip}
\usepackage{textcomp}
\usepackage{subcaption}
\usepackage[noadjust]{cite}
\usepackage{makecell}

\begin{document}

\title{HotFuzz: Discovering Algorithmic Denial-of-Service Vulnerabilities
    Through Guided Micro-Fuzzing}

\author{\IEEEauthorblockN{William Blair\IEEEauthorrefmark{2},
Andrea Mambretti\IEEEauthorrefmark{1},
Sajjad Arshad\IEEEauthorrefmark{3},
Michael Weissbacher\IEEEauthorrefmark{1},\\
William Robertson\IEEEauthorrefmark{1},
Engin Kirda\IEEEauthorrefmark{1}, and
Manuel Egele\IEEEauthorrefmark{2}
}
\IEEEauthorblockA{\IEEEauthorrefmark{2}Boston University}
\IEEEauthorblockA{\IEEEauthorrefmark{1}\IEEEauthorrefmark{3}Northeastern University}
\IEEEauthorblockA{\IEEEauthorrefmark{2}\{wdblair, megele\}@bu.edu, \IEEEauthorrefmark{1}\{mbr, mw, wkr, ek\}@ccs.neu.edu, \IEEEauthorrefmark{3}arshad@iseclab.org}
}

\IEEEoverridecommandlockouts
\makeatletter\def\@IEEEpubidpullup{6.5\baselineskip}\makeatother
\IEEEpubid{\parbox{\columnwidth}{
    Network and Distributed Systems Security (NDSS) Symposium 2020\\
    23-26 February 2020, San Diego, CA, USA\\
    ISBN 1-891562-61-4\\
    https://dx.doi.org/10.14722/ndss.2020.24415\\
    www.ndss-symposium.org
}
\hspace{\columnsep}\makebox[\columnwidth]{}}

\maketitle

\begin{abstract}
\javacount devices run Java and many of them are connected to the Internet. As
this ecosystem continues to grow, it remains an important task to discover any
unknown security threats these devices face. Fuzz testing repeatedly runs
software on random inputs in order to trigger unexpected program behaviors,
such as crashes or timeouts,  and has historically revealed serious security
vulnerabilities. Contemporary fuzz testing techniques focus on identifying
memory corruption vulnerabilities that allow adversaries to achieve either
remote code execution or information disclosure. Meanwhile, Algorithmic
Complexity (AC) vulnerabilities, which are a common attack vector for
denial-of-service attacks, remain an understudied threat.

In this paper, we present \thesystem, a framework for automatically discovering
AC vulnerabilities in Java libraries. \thesystem uses micro-fuzzing, a genetic
algorithm that evolves arbitrary Java objects in order to trigger the
worst-case performance for a method under test. We define Small Recursive
Instantiation (\sri) as a technique to derive seed inputs represented as Java
objects to micro-fuzzing. After micro-fuzzing, \thesystem synthesizes test
cases that triggered AC vulnerabilities into Java programs and monitors their
execution in order to reproduce vulnerabilities outside the fuzzing framework.
\thesystem outputs those programs that exhibit high CPU utilization as
witnesses for AC vulnerabilities in a Java library.

We evaluate \thesystem over the Java Runtime Environment (JRE), the 100 most
popular Java libraries on Maven, and challenges contained in the \stac program.
We evaluate \sri's effectiveness by comparing the performance of micro-fuzzing
with \sri, measured by the number of AC vulnerabilities detected, to simply
using empty values as seed inputs. In this evaluation, we verified known AC
vulnerabilities, discovered previously unknown AC vulnerabilities that we
responsibly reported to vendors, and received confirmation from both IBM and
Oracle. Our results demonstrate that micro-fuzzing finds AC vulnerabilities in
real-world software, and that micro-fuzzing with \sri-derived seed inputs
outperforms using empty values.\end{abstract}

\section{Introduction}
\label{sec:introduction}

Software continues to be plagued by vulnerabilities that allow attackers to
violate basic software security properties. These vulnerabilities take myriad
forms, for instance failures to enforce memory safety that can lead to arbitrary
code execution (integrity violations) or failures to prevent sensitive data from
being released to unauthorized principals (confidentiality violations).  The
third traditional security property, availability, is not exempt from this issue.
However, denial-of-service (DoS) as a vulnerability class tends to
be viewed as simplistic, noisy, and easy (in principle) to defend against.

This view, however, is simplistic, as availability vulnerabilities and exploits
against them can take sophisticated forms.
Algorithmic Complexity (AC) vulnerabilities are one such form, where a small
adversarial input induces worst-case\footnote{Strictly speaking, it is
sufficient for attacks to cause bad behavior, it need not be
``worst-case''.} behavior in the processing of that input, resulting in a denial
of service.

While in the textbook example against a hash table an adversary inserts values
with colliding keys to degrade the complexity of lookup operations from an
expected \(O(1)\) to \(O(n)\), the category of AC vulnerabilities is by no
means hypothetical. Recent examples of algorithmic complexity vulnerabilities
include denial of service issues in Go's elliptic curve cryptography
implementation~\cite{golangcve}, an AC vulnerability that manifests through
amplification of API requests against Netflix' internal infrastructure
triggered by external requests~\cite{netflix}, and a denial of service
vulnerability in the Linux kernel's handling of TCP packets~\cite{linuxcve}.
The vulnerability in the Linux kernel was considered serious enough that it was
embargoed until OS vendors and large Linux users such as cloud providers and
content delivery networks could develop and deploy patches.
While these particular vulnerabilities involved unintended CPU time complexity,
AC vulnerabilities can also manifest in the spatial domain for resources such as
memory, storage, or network bandwidth.

While discovering AC vulnerabilities is notoriously challenging, program
analysis seems like a natural basis for developing solutions to tackle such
issues.
In fact, prior research has started to explore program analysis techniques for
finding AC vulnerabilities in software. Most of this work is based on manual or
static analysis that scales to real world code bases, but focuses on detecting
known sources of AC vulnerabilities, such as triggering worst case performance
of commonly used data structures~\cite{crosby_denial_2003}, regular
expression engines~\cite{kirrage13:regex_dos,StaicuP18,wustholz_static_2017},
or serialization APIs~\cite{dietrich_evil_2017}.

Fuzz testing, where a fuzzer feeds random input to a program under test until
the program either crashes or times out, has historically revealed serious bugs
that permit Remote Code-Execution (RCE) exploits in widely used software such
as operating system kernels, mobile devices, and web browsers. Recent work has
adapted existing state-of-the-art fuzz testers such as AFL~\cite{afl} and
libFuzzer~\cite{libFuzzer} to automatically slow down programs with known
performance problems. These approaches include favoring inputs that maximize
the length of an input's execution in a program's Control Flow Graph
(CFG)~\cite{PetsiosZKJ17}, incorporating multi-dimensional feedback that
provides AFL with more visibility into the portions of the CFG each test case
executes the most~\cite{LemieuxPSS18}, and augmenting AFL with symbolic
execution to maximize a Java program's resource consumption~\cite{NollerKP18}.
These recent advances demonstrate that modern fuzzers can automatically slow
down programs such as sorting routines, hash table operations, and common Unix
utilities.

These recent developments present exciting new directions for fuzz testing
beyond detecting memory corruption bugs. However, these approaches do not
reconcile the traditional fuzzing objective function of maximizing code
coverage (breadth) with the opposing goal of maximizing a given program or
individual method's runtime (depth). Indeed, these tools are evaluated by the
slowdown they can achieve for a given program, as opposed to the amount of code
they successfully cover.  Achieving high code coverage on any program under
test is a notoriously difficult task because common program patterns like
comparing input to magic values or checksum tests are difficult to bypass using
fuzzing alone, although program transformation tricks like splitting each
comparison into a series of one byte comparisons~\cite{laf} or simply removing
them from the program~\cite{PengSP18} can improve coverage.  Augmenting fuzzing
with advanced techniques like taint analysis~\cite{rawat_vuzzer:_2017} or
symbolic execution~\cite{stephens_driller:_2016,munch18} helps overcome these
fuzzing roadblocks, and RedQueen~\cite{redqueen19} showed how advanced tracing
hardware can emulate these more heavyweight techniques by providing a fuzzer
with enough information to establish correspondence between program inputs and
internal program state.  Prior work has successfully shown fuzz testing can
reproduce known AC vulnerabilities in software, and research continues to
produce innovative ways to maximize code coverage. What is missing in fuzzing
for AC vulnerabilities are techniques to automatically sanitize a program's
execution for AC vulnerabilities, analogous to how modern fuzzers rely on
sanitizers to detect memory corruption bugs~\cite{asan12}.  Current fuzzing
approaches in general lack the ability to automatically fuzz programs at the
method level without the need for manually defined test harnesses.

This paper proposes \emph{micro-fuzzing} (a concept analogous to
micro-execution~\cite{godefroid2014}) as a novel technique to automatically
construct test harnesses that allow a fuzzer to invoke methods and sanitize
their execution for AC vulnerabilities.
Both AFL and libFuzzer can fuzz individual methods, but only after an analyst
manually defines a test harness that transforms a flat bitmap into the types
required to call a method. For AFL this involves defining a C program that
reads the bitmap from standard input, whereas libFuzzer passes the bitmap to a
specific function that it expects will call the method under test with the appropriate
types derived from the bitmap.

In contrast, micro-fuzzing takes whole programs or libraries as input and
attempts to automatically construct a test harness for every function contained
in the input.
Observe that this approach is analogous to
micro-execution~\cite{godefroid2014}, which executes arbitrary machine code by
using a virtual machine as a test harness that provides state on-demand in
order to run the code under test.
To this end, micro-fuzzing constructs test harnesses represented as function
inputs, directly invokes functions on those inputs, and measures the amount of
resources each input consumes using model specific registers available on the
host machine. This alleviates the need to define test harnesses manually, and
supports fuzzing whole programs and libraries by considering every function
within them as a possible entrypoint. Furthermore, we sanitize every function's
execution so that once its observed runtime crosses a configured threshold, we
kill the micro-fuzzing process and highlight the function as vulnerable. This
sanitization highlights functions with potential AC vulnerabilities out of all
the functions micro-fuzzing automatically executes, as opposed to measuring a
fuzzer's ability to automatically slow-down individual programs or functions.

We implement micro-fuzzing for Java programs in \thesystem, which uses a
genetic algorithm to evolve method inputs with the goal to maximize method
execution time. Java provides an ideal platform for evaluating micro-fuzzing
because of its wide use across different domains in industry and the JVM's
support for introspection allows \thesystem to automatically generate test
harnesses, represented as valid Java objects, for individual methods
dynamically at runtime.
To generate initial populations of inputs, we devise two different strategies.
The Identity Value Instantiation (\ivi) strategy creates inputs by assigning
each actual parameter the identity element of the parameter's domain (e.g., 0
for numeric types or ``'' for strings). In contrast, Small Recursive
Instantiation (\sri) assigns parameters small values chosen at random from the
parameter's domain. We use \ivi for the sole purpose of providing a baseline
for measuring the effectiveness of using \sri to generate seed inputs for
micro-fuzzing, based on recent recommendations for evaluating new fuzz testing
techniques~\cite{KleesRCW018}.

Irrespective of how inputs are instantiated, \thesystem leverages the \eyevm,
an instrumented JVM that provides run-time measurements at method-level
granularity. If micro-fuzzing creates an input that causes the method under
test's execution time to exceed a threshold, \thesystem marks the method as
potentially vulnerable to an AC attack. To validate potential AC
vulnerabilities, \thesystem synthesizes Java programs that invoke flagged
methods on the suspect inputs and monitors their end-to-end execution in an
unmodified JVM that mirrors a production environment. Those programs that
exceed a timeout are included in \thesystem's output corpus.  Every program
contained in the output corpus represents a witness of a potential AC
vulnerability in the library under test that a human operator can either
confirm or reject. Sanitizing method execution for AC vulnerabilities based on
a threshold mimics the sanitizers used by modern fuzzers that kill a process
whenever an integrity violation occurs at runtime, but it also introduces false
positives into our results given that it is difficult to configure a proper
timeout that detects only true positives. In our evaluation, we show that the
number of bugs detected by our sanitizer is concise enough to permit manual
analysis of the results.

We evaluate \thesystem by micro-fuzzing the Java Runtime Environment (JRE),
challenges provided by the \stac program, and the 100 most popular libraries
available on Maven, a popular repository for hosting Java program dependencies.
We identify \solvedchallenges intentional (in STAC) and \totalbugs
unintentional (in the JRE and Maven libraries) AC vulnerabilities.

In summary, this paper makes the following contributions:

\begin{itemize}
  \item We introduce micro-fuzzing as a novel and efficient technique for
    identifying AC vulnerabilities in Java programs (see
    Section~\ref{sec:microfuzzing}).
  \item We devise two strategies (\ivi and \sri) to generate seed
    inputs for micro-fuzzing (see Section~\ref{sec:sri}).
  \item We propose the combination of \ivi and \sri with micro-fuzzing to
    detect AC vulnerabilities in Java programs.
  \item We design and evaluate \thesystem, an implementation of our
   micro-fuzzing approach, on the Java Runtime Environment (JRE), challenges
   developed during the DARPA STAC program, and the 100 most popular libraries
   available on Maven.
    Our evaluation results yield previously unknown AC vulnerabilities in
    real-world software, including \zerodaynum in the JRE, \mavenzerodaynum
    across \vulnmavenlibraries Maven libraries, including the widely used
    org.json library, ``solve'' \solvedchallenges challenges from the
    STAC program, and include confirmations from IBM and Oracle. In addition,
    micro-fuzzing with \sri-derived seed inputs outperforms \ivi-derived seed
    inputs, measured by the number of AC witnesses detected
    (see Section~\ref{sec:evaluation}).
\end{itemize}

\section{Background and Threat Model}
\label{sec:motivation}

In this section, we briefly describe Algorithmic Complexity (AC)
vulnerabilities, different approaches that detect such vulnerabilities, the
threat model we assume, and the high-level design goals of this work.

\subsection{AC Vulnerabilities}

AC vulnerabilities arise in programs whenever an adversary can provide inputs
that cause the program to exceed desired (or required) bounds in either the
spatial or temporal domains.  One can define an AC vulnerability in terms of
asymptotic complexity (e.g., an input of size \(n\) causes a method to store
\(O(n^3)\) bytes to the filesystem instead of the expected \(O(n)\)), in terms
of a concrete function of the input (e.g., an input of size \(n\) causes a
method to exceed the intended maximum \(150n\) seconds of wall clock execution
time), or in other more qualitative senses (e.g., ``the program hangs for
several minutes'').  However, in each case there is a definition, explicit or
otherwise, of what constitutes an acceptable resource consumption threshold.

In this work, we assume an explicit definition of this threshold independent of
a given program under analysis and rely on domain knowledge and manual
filtering of AC witnesses in order to label those that should be considered as
true vulnerabilities.  We believe that this is a realistic assumption and
pragmatic method for vulnerability identification that avoids pitfalls
resulting from attempting to automatically understand intended resource
consumption bounds, or from focusing exclusively on asymptotic complexity when
in practice, as the old adage goes, ``constants matter.'' We define an AC
witness to be any input that causes a specific method under test's resource
consumption to exceed a configured threshold. We consider any method that has
an AC witness to contain an AC vulnerability.

We recognize that this definition of an AC vulnerability based on observing a
method's resource consumption exceeding some threshold will inevitably cause
some false positives, since the chosen threshold may not be appropriate for a
given method under test.  Section~\ref{sec:system_overview} presents a strategy
for minimizing  false positives by automatically reproducing AC vulnerabilities
in a production environment outside our fuzzing framework. This step may fail
to remove all false positives, and in our evaluation given in
Section~\ref{sec:evaluation} we show that the output of this validation stage
is concise enough to allow an analyst to manually triage the results. Since we
make no assumption about the methods we test in our analysis, we believe
observing output that consists of less than three hundred test cases is
reasonable for a human analyst.

\subsection{AC Detection}

Software vulnerability detection in general can be roughly categorized as a
static analysis, dynamic testing, or some combination of the two.
Static analysis has been proposed to analyze a given piece of code for its worst
case execution time behavior. While finding an upper bound to program execution
time is certainly valuable, conservative approximations in static analysis
systems commonly result in a high number of false positives. Furthermore, even
manual interpretation of static analysis results in this domain can be
challenging as it is often unclear whether a large worst-case execution time
results from a property of the code or rather the approximation in the analysis.
Additionally, static analyses for timing analysis commonly work best for well
structured code that is written with such analysis in mind (e.g., code in a
real-time operating system).
The real-world generic code bases in our focus (e.g., the Java Runtime
Environment), have not been engineered with such a focus and quickly reach the
scalability limits of static timing analyzers.

Dynamic testing, in particular fuzz testing, has emerged as a particularly
effective vulnerability detection approach that runs continuously in
parallel with the software development lifecycle~\cite{msfuzz,ossfuzz}.
State of the art fuzzers detect bugs by automatically executing a program under
test instrumented with sanitizers until the program either crashes or times
out. A sanitized program crashes immediately after it violates an invariant
enforced by the sanitizer, such as writing past the boundary of a buffer
located on the stack or reading from previously freed memory. Once a fuzzer
generates a test case that crashes a given sanitized program under test, the
test case is a witness for a memory corruption bug in the original program.
Since memory corruption bugs may be extended into exploits that achieve Remote
Code Execution or Information Disclosure, fuzzers offer an effective and
automated approach to software vulnerability detection.
When source code is not available, a fuzzer can still attempt to crash the program under
test in either an emulated or virtualized environment.

Fuzz testing's utility for detecting memory corruption bugs in programs is
well known, and current research explores how to maximize both the amount of code a
fuzzer can execute and the number of bugs a fuzzer can find. Unfortunately,
defining a sanitizer that crashes a process after an AC vulnerability occurs
is not as straightforward as detecting memory integrity violations.  This is in
part because what constitutes an AC vulnerability heavily depends on the
program's domain.  For example, a test case that slows down a program by
several milliseconds may be considered an AC vulnerability for a low latency
financial trading application and benign for a web service that processes
requests asynchronously.

In this work, we propose a sanitizer in \thesystem that kills a process after a
method's runtime exceeds a configured threshold. Like sanitizers for memory
corruption bugs, this allows us to save only those test cases that exhibit
problematic behavior. The drawback is that we do not have absolute certainty
that our test cases are actual bugs in the original program and risk
highlighting test cases as false positives. Building a fuzzing analysis that
does not introduce any false positives is notoriously difficult, and fuzzers
that detect memory corruption bugs are not immune to this problem. For
example, Aschermann~\textit{et al.}~\cite{redqueen19} point out that previous
evaluations erroneously report crashing inputs that exhaust the fuzzer's
available memory as bugs in the original program under test. Furthermore,
sanitizers point out many different sources of bugs including stack
based overflows, use after free, use after return, and heap based overflows.
While the presence of any of these bugs is problematic, triaging is still
required to understand the problem given in a test case.

\subsection{Fuzzing AC}

SlowFuzz~\cite{PetsiosZKJ17} and PerfFuzz~\cite{LemieuxPSS18}, adapt two state
of the art fuzzers, libFuzzer and AFL, respectively, and demonstrate the
capability to automatically slow down individual programs or methods
implemented in C/C++. Parallel developments also showed frameworks built on top
of AFL can successfully slow down programs in interpreted languages as
well~\cite{NollerKP18}.

\thesystem departs from these previous works by automatically creating
test harnesses during micro-fuzzing, and sanitizing method execution for AC
vulnerabilities.
In contrast to these tools, \thesystem does not require an analyst to manually
define a test harness in order to fuzz individual methods contained in a
library.  This key feature differentiates micro-fuzzing found in \thesystem
from how AFL or libFuzzer fuzz individual methods.  Since AFL and LibFuzzer
only consider test cases consisting of flat bitmaps, one can fuzz an individual
method with AFL by defining a test harness that transforms a bitmap read from
\texttt{stdin} into function inputs, and with libFuzzer an analyst implements a
\texttt{C} function that takes the test case as input and must transform it into
the types needed to invoke a function. Observe that this must be done manually,
whereas \thesystem examines the type signature of the method under test and
attempts to generate the test harness automatically. To reproduce our
evaluation using existing tools, we would need to manually define approximately
400,000 individual test harnesses for all the artifacts contained in our
evaluation.

SlowFuzz and PerfFuzz both explore how fuzzers can automatically slow down
individual programs. Understanding what techniques work best to slow down code
is necessary to understand how to design a fuzzer to detect AC vulnerabilities.
SlowFuzz observed that using the number of executed instructions as a test
case's fitness in libFuzzer's genetic algorithm can be used to slow down code
with known performance problems, such as sorting routines and hash table
implementations. PerfFuzz went a step further and showed how incorporating a
performance map that tracks the most visited edges in a program's CFG can help
a fuzzer further slow down programs.

These approaches take important steps needed to understand what techniques
allow fuzzers to automatically slow down arbitrary code in order to spot AC
vulnerabilities in programs. At the same time, they lack three important
properties for being used to detect unknown AC vulnerabilities. First, they
require manually defined test harnesses in order to fuzz individual functions.
Second, these fuzzing engines only consider flat bitmaps as input to the
programs under test, and miss the opportunity to evolve the high level classes
of the function's domain in the fuzzer's genetic algorithm. Finally, these
tools are meant to understand what techniques successfully slow down code the
most, and do not provide a method for sanitizing method execution for AC
vulnerabilities and presenting these results to a human analyst.

\subsection{Optimization}
The goal of identifying AC vulnerabilities boils down to a simple to posit yet
challenging to answer optimization question.
``What are concrete input values that make a given method under test consume the
most resources?''
One possible approach to tackle such optimization problems is with the help of
genetic algorithms.
A genetic algorithm emulates the process of evolution to derive approximations
for a given optimization problem.
To this end, a genetic algorithm will start with an initial population of
individuals and over the duration of multiple generations repeatedly perform
three essential steps:
\begin{inparaenum}[i)]
    \item Mutation,
    \item Crossover, and
    \item Selection.
\end{inparaenum}
In each generation, a small number of individuals in the population may undergo
mutation. Furthermore, each generation will see a large number of crossover
events where two individuals combine to form offspring. Finally, individuals in
the resulting population get evaluated for their fitness, and the individuals
with the highest fitness are selected to form the population for the next
generation.
The algorithm stops after either a fixed number of generations, or when the
overall fitness of subsequent populations no longer improves.
In our scenario where we seek to identify AC vulnerabilities in Java methods,
individuals correspond to the actual parameter values that are passed to a
method under test.
Furthermore, assessing fitness of a given individual can be accomplished by
measuring the method's resource consumption while processing the individual (see
Section~\ref{sec:instrumented_jvm}).
While mutation and crossover are straightforward to define on populations whose
individuals can be represented as sequences of binary data, the individuals in
our setting are tuples of well-formed Java objects. As such, mutation and
crossover operators must work on arbitrary Java classes, as opposed to flat
binary data (see Section~\ref{sec:fuzzer}).

\subsection{Threat Model}

In this work, we assume the following adversarial capabilities.  An attacker
either has access to the source code of a targeted program and its
dependencies, or a compiled artifact that can be tested offline.  Using this
code, the attacker can employ arbitrary techniques to discover AC
vulnerabilities exposed by the program, either in the program itself or by any
library functionality invoked by the program. Furthermore, we assume that these
vulnerabilities can be triggered by untrusted input.

An adversary can achieve DoS attacks on programs and services that utilize
vulnerable libraries by taking the information they learn about a library
through offline testing and developing exploits that trigger the AC
vulnerabilities contained in library methods used by a victim program. For
example, an adversary could take the test cases produced by our evaluation (see
Section~\ref{sec:evaluation}) and attempt to reproduce their behavior on
programs that utilize the methods. Determining whether an adversary can
transform these test cases into working AC exploits on victim programs is
outside the scope of this work.

\subsection{Design Goals}

The goal of our work is to discover AC vulnerabilities in Java code so that
they can be patched before attackers have the opportunity to exploit them.  In
particular, we aim for an analysis that is automated and efficient such that it
can run continuously in parallel with the software development lifecycle on
production artifacts. This gives developers insight into potential
vulnerabilities hiding in their applications without altering their development
workflow.

\section{\thesystem Overview}
\label{sec:system_overview}

\begin{figure*}[t]
    \centering
    \includegraphics[width=0.9\textwidth]{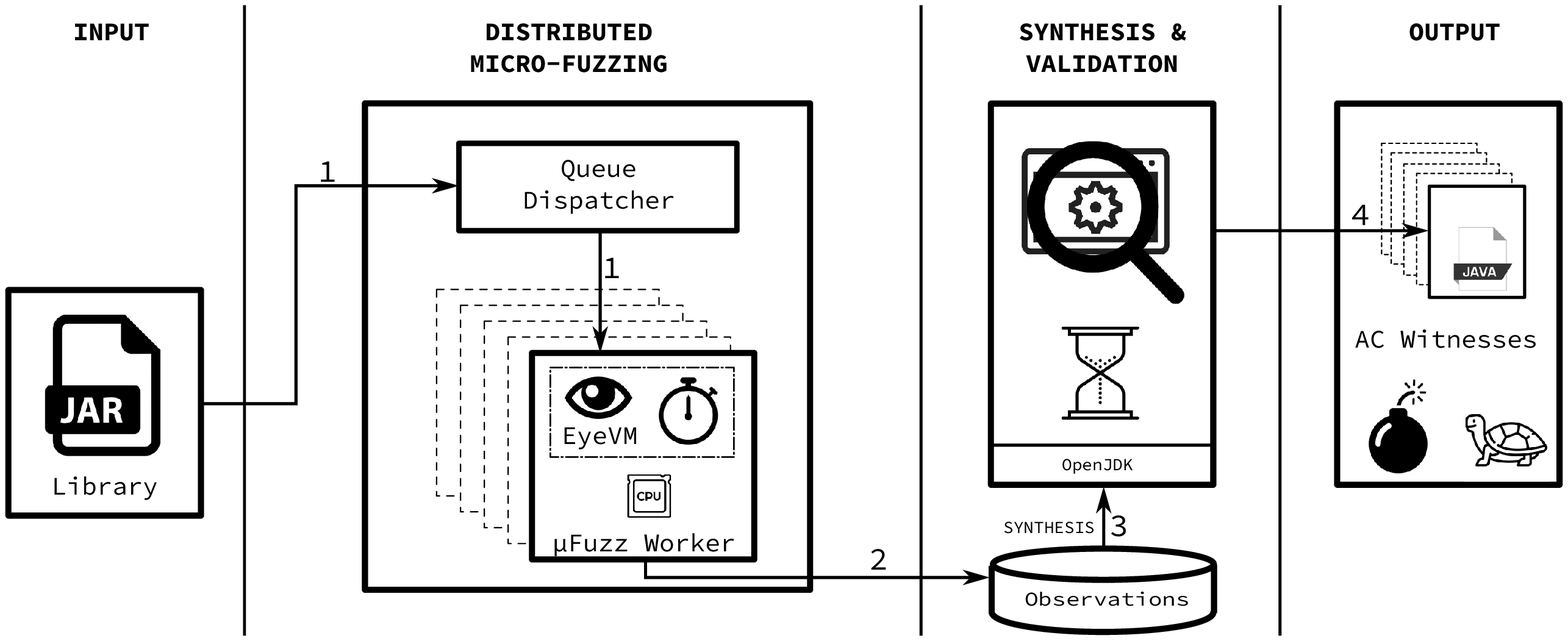}
    \caption{Architectural overview of the \thesystem testing procedure. In the first phase, individual \fuzzer instances micro-fuzz each method comprising a library under test. Resource consumption is maximized using genetic optimization over inputs seeded using either \ivi or \sri. In the second phase, test cases flagged as potential AC vulnerabilities by the first phase are synthesized into Java programs. These programs are executed in an unmodified JVM in order to replicate the abnormal resource consumption observed in the first phase.  Programs that fail to do so are rejected as false positives. \thesystem reports those programs that pass the Synthesis and Validation stage as AC vulnerability witnesses to a human analyst.}
    \label{fig:architecture}
\end{figure*}

\thesystem adopts a dynamic testing approach to detecting AC vulnerabilities, where the testing procedure consists of two phases:
\begin{inparaenum}[\itshape (i)\upshape]
    \item micro-fuzzing, and
    \item witness synthesis and validation.
\end{inparaenum}
In the first phase, a Java library under test is submitted for
\emph{micro-fuzzing}, a novel approach to scale AC vulnerability detection.  In
this process, the library is decomposed into individual methods, where each
method is considered a distinct entrypoint for testing by a \emph{\fuzzer}
instance.  As opposed to traditional fuzzing, where the goal is to provide
inputs that crash a program under test, here each \fuzzer instance attempts to
maximize the resource consumption of individual methods under test using genetic
optimization over the method's inputs.  To that end, seed inputs for each method under
test are generated using one of two instantiation strategies: \emph{Identity
Value Instantiation} (\ivi) and \emph{Small Recursive Instantiation} (\sri).
Method-level resource consumption when executed on these inputs is measured
using a specially-instrumented Java virtual machine we call the \emph{EyeVM}.
If optimization eventually produces an execution that is measured to exceed a
pre-defined threshold, then that test case is forwarded to the second phase of
the testing procedure.

Differences between the micro-fuzzing and realistic execution environments can
lead to false positives. The purpose of the second phase is to validate whether
test cases found during micro-fuzzing represent actual vulnerabilities when
executed in a real Java run-time environment, and therefore reduce the number
of false positives in our final results.  This validation is achieved through
\emph{witness synthesis} where, for each test case discovered by the first
phase, a program is generated that invokes the method under test with the
associated inputs that produce abnormal resource usage.  If the behavior with
respect to resource utilization that was observed during micro-fuzzing is
replicated, then the synthesized test case is flagged as a witness of the
vulnerability that can then be examined by a human analyst.  Otherwise, we
discard the synthesized test case as a false positive.

Figure~\ref{fig:architecture} depicts a graphical overview of the two phases.  In the following, we motivate and describe the design of each component of the testing procedure in detail.

\subsection{Micro-Fuzzing}
\label{sec:microfuzzing}

Micro-fuzzing represents a drastically different approach to vulnerability detection than traditional automated whole-program fuzzing.  In the latter case, inputs are generated for an entire program either randomly, through mutation of seed inputs, or incorporating feedback from introspection on execution.  Whole-program fuzzing has the significant benefit that any abnormal behavior---i.e., crashes---that is observed should be considered as a real bug as by definition all the constraints on the execution path that terminates in the bug are satisfied (up to the determinism of the execution).  However, whole-program fuzzing also has the well-known drawback that full coverage of the test artifact is difficult to achieve.  Thus, an important measure of a traditional fuzzer's efficacy is its ability to efficiently cover paths in a test artifact.

Micro-fuzzing strikes a different trade-off between coverage and path satisfiability.  Inspired by the concept of micro-execution~\cite{godefroid2014}, micro-fuzzing constructs realistic intermediate program states, defined as Java objects, and directly executes individual methods on these states. Thus, we can cover all methods by simply enumerating all the methods that comprise a test artifact, while the difficulty lies instead in ensuring that constructed states used as method inputs are feasible in practice.\footnote{We note that in the traditional fuzzing case, a similar problem exists in that while crashes indicate the presence of an availability vulnerability, they do not necessarily represent exploitable opportunities for control-flow hijacking.}  In our problem setting, where we aim to preemptively warn developers against insecure usage of AC-vulnerable methods or conservatively defend against powerful adversaries, we believe micro-fuzzing represents an interesting and useful point in the design space that complements whole program fuzzing approaches. In this work, we consider the program's state as the inputs given to the methods we micro-fuzz. Modeling implicit parameters, such as files, static variables, or environment variables are outside the scope of this work.

A second major departure from traditional fuzzing is the criteria used to identify vulnerabilities.  Typical fuzzers use abnormal termination as a signal that a vulnerability might have been found.  In our case, vulnerabilities are represented not by crashes but rather by excessive resource consumption.  Thus, coverage is not the sole metric that must be maximized in our case.  Instead, \thesystem must balance between maximizing a method's resource utilization \emph{in addition} to coverage.  Conceptually speaking, implementing resource measurement is a straightforward matter of adding methods to the existing Reflection API in Java that toggles resource usage recording and associates measurements with Java methods.  In practice, this involves non-trivial engineering, the details of which we present in Section~\ref{sec:implementation}.  In the following, we describe how \thesystem optimizes resource consumption during micro-fuzzing given dynamic measurements provided by the \eyevm, our instrumented JVM that provides run-time measurements at method-level granularity.

\subsubsection{Resource Consumption Optimization}
\label{sec:fuzzer}

\thesystem's fuzzing component, called \fuzzer, is responsible for optimizing the resource consumption of methods under test.  To do so, \fuzzer uses genetic optimization to evolve an initial set of seed inputs over multiple generations until it detects abnormal resource consumption. Traditional fuzzers use evolutionary algorithms extensively, but in this work we present a genetic optimization approach to fuzzing that departs from prior work in two important ways. First, as already discussed, traditional fuzzers optimize an objective function that solely considers path coverage (or some proxy thereof), whereas in our setting we are concerned in addition with resource consumption. Prior work for detecting AC vulnerabilities through fuzz testing either record resource consumption using a combination of program instrumentation, CPU utilization, or counting executed instructions. In contrast, we record resource consumption using an altered execution environment (the \eyevm) and require no modification to the library under test.  Second, traditional fuzzers treat inputs as flat bitmaps when genetic optimization (as opposed to more general mutation) is applied.  Recall that genetic algorithms require defining crossover and mutation operators on members of the population of inputs.  New generations are created by performing crossover between members in prior generations.  Additionally, in each generation, some random subset of the population undergoes mutation with small probability.  Since \fuzzer operates on Java values rather than flat bitmaps, we must define new crossover and mutation operators specific to this domain as bitmap-specific operators do not directly translate to arbitrary Java values, which can belong to arbitrary Java classes.

\paragraph{Java Value Crossover}
\label{sec:crossover_input}
Genetic algorithms create new members of a population by ``crossing'' existing members.  When individual inputs are represented as bitmaps, a standard approach is single-point crossover: a single offset into two bitmaps is selected at random, and two new bitmaps are produced by exchanging the content to the right of the offset from both parents.  Single-point crossover does not directly apply to inputs comprised of Java objects, but can be adapted in the following way.  Let \(X_0, X_1\) represent two existing inputs from the overall population and \((x_0, x_1)_0 = x_0\) and \((x_0, x_1)_1 = x_1\).  To produce two new inputs, perform single-point crossover for each corresponding pair of values \((x_0, x_1) \in (X_0, X_1)\) using
\[ \scalemath{0.9}{
    (x_0', x_1') =
\begin{cases}
    C(x_0, x_1) & \quad \text{if } (x_0, x_1) \text{ are primitives} \\
    (C_L(x_0, x_1), C_R(x_0, x_1)) & \quad \text{if } (x_0, x_1) \text{ are objects.} \\
\end{cases}
}
\]

Here, \(C\) performs one-point crossover directly on primitive values and produces the offspring as a pair. When \(x_0\) and \(x_1\) are objects, \(C_L\) and \(C_R\) recursively perform cross-over on every member attribute in \((x_0, x_1)\) and select the left and right offspring, respectively.
For example, consider a simple Java class \texttt{List} that implements a singly linked list. The \texttt{List} class consists of an \texttt{integer} attribute \texttt{hd} and a \texttt{List} attribute \texttt{tl}. Crossing an instance of \texttt{List} $\vec{x}$ with another instance $\vec{y}$ constructs two new lists $\vec{x'}$ and $\vec{y'}$ given by

\begin{align*}
\vec{x'} & = C_L(\vec{x}, \vec{y}) = \left(hd := C(\vec{x}.hd, \vec{y}.hd)_0, tl := C_L(\vec{x}.tl, \vec{y}.tl)\right) \\
\vec{y'} & = C_R(\vec{x}, \vec{y}) = \left(hd := C(\vec{x}.hd, \vec{y}.hd)_1, tl := C_R(\vec{x}.tl, \vec{y}.tl)\right)
\end{align*}

In this example we show how \thesystem crosses over a List that holds Integers,
but if the type of value stored in the \texttt{hd} attribute were a complex
class \(T\), the crossover operator would recursively apply crossover to
every attribute stored in \(T\).

\paragraph{Java Value Mutation}
\label{sec:mutating_input}
Mutation operators for traditional fuzzers rely on heuristics to derive new generations, mutating members of the existing population through random or semi-controlled bit flips.  In contrast, micro-fuzzing requires mutating arbitrary Java values, and thus bitmap-specific techniques do not directly apply.

Instead, \fuzzer mutates Java objects using the following procedure.  For a given Java object \(x\) with attributes \(\{a_0, a_1, \ldots, a_n\}\), choose one of its attributes \(a_i\) uniformly at random.  Then we define the mutation operator \(M\) as
\[
a_i' =
\begin{cases}
    M_{\textsf{flip\_bit}}(a_i) & \quad \text{if } a_i \text{ is a numeric value,} \\
    M_{\textsf{insert\_char}}(a_i) & \quad \text{if } a_i \text{ is a string or array value,} \\
    M_{\textsf{delete\_char}}(a_i) & \quad \text{if } a_i \text{ is a string or array value,} \\
    M_{\textsf{replace\_char}}(a_i) & \quad \text{if } a_i \text{ is a string or array value,} \\
    M_{\textsf{swap\_chars}}(a_i) & \quad \text{if } a_i \text{ is a string or array value,} \\
    M_{\textsf{mutate\_attr}}(a_i) & \quad \text{if } a_i \text{ is an object.} \\
\end{cases}
\]
Each mutation sub-operator above operates on the attribute $a_i$ chosen from
the object \(x\). For example, \(M_{\textsf{flip\_bit}}\) selects a bit at random
in a numeric element and flips it, while \(M_{\textsf{swap\_chars}}\) randomly
selects two elements of a string or array and swaps them. In our current
implementation, we only consider arrays of primitive types. The other
sub-operators are defined in an intuitively similar manner.

When an attribute is a class, as opposed to a primitive type or a string or
array, mutation utilizes the \(M_\textsf{mutate\_attr}\) operator.
\(M_\textsf{mutate\_attr}\) recursively applies the mutation operator \(M\) to
the chosen attribute \(a_i\) when \(a_i\) is an object. After we obtain the
mutated attribute \(a_i'\), we produce the mutated object \(x'\) by replacing
\(a_i\) with \(a_i'\) in \(x\).

\subsubsection{Seed Generation}
\label{sec:seed_generation}

Given suitable crossover and mutation operators, all that remains to apply standard genetic optimization is the definition of a procedure to generate seed inputs.  We define two such procedures that we describe below: Identity Value Instantiation (\ivi), and Small Recursive Instantiation (\sri).

\paragraph{Identity Value Instantiation}
\label{sec:ivi}
Recent work has proposed guidelines for evaluating new fuzz testing
techniques~\cite{KleesRCW018}.  One of these guidelines is to compare any
proposed strategy for constructing seed inputs for fuzz testing with ``empty''
seed inputs. Intuitively, empty seed inputs represent the simplest possible
seed selection strategy. Since empty bitmaps do not directly translate to our
input domain, we define \ivi as an equivalent strategy for Java values.  The
term ``identity value'' is derived from the definition of an identity element
for an additive group.

In particular, \ivi is defined as
\[\scalemath{0.9}{
    I(T) =
\begin{cases}
    0 & \quad \text{if } T \text{ is a numeric type}, \\
    false & \quad \text{if } T \text{ is a boolean}, \\
    "" & \quad \text{if } T \text{ is a string}, \\
    \{\} & \quad \text{if } T \text{ is an array}, \\
    T_{\textsf{random}}\left(I(T_0), \ldots, I(T_n)\right) & \quad \text{if } T \text{ is a class.} \\
\end{cases}
}\]
That is, \(I(T)\) selects the identity element for all primitive types, while for classes \(I\) is recursively applied to all parameter types \(T_i\) of a randomly selected constructor for \(T\).  Thus, for a given method under test \(M\), \(I(M)\) is defined as \(I\) applied to each of \(M\)'s parameter types.

\paragraph{Small Recursive Instantiation}
\label{sec:sri}
In addition to \ivi, we define a complementary seed input generation procedure called Small Recursive Instantiation (\sri).  In contrast to \ivi, \sri generates random values for each method parameter.  However, experience dictates that selecting uniformly random values from the entire range of possible values for a given type is not the most productive approach to input generation. For example, starting with large random numbers as seed inputs may waste time executing benign methods that simply allocate large empty data structures like Lists or Sets. For example, creating a List with the \texttt{ArrayList(int capacity)} constructor and passing it an initial capacity of \texttt{1<<30} takes over 1 second and requires over \texttt{4GB} of RAM. For this reason, we configure \sri with a spread parameter \(\alpha\) that limits the range of values from which \sri will sample.  Thus, \sri is defined as
\[\scalemath{0.85} {
    S(T, \alpha) =
\begin{cases}
    R_{\textsf{num}}(-\alpha, \alpha) & \quad \text{if } T \text{ is a numeric type}, \\
    \left\{R_{\textsf{char}}\right\}^{R_{\textsf{num}}(0, \alpha)} & \quad \text{if } T \text{ is a string}, \\
    \left\{S(T, \alpha)\right\}^{R_{\textsf{num}}(0, \alpha)} & \quad \text{if } T \text{ is an array}, \\
    T_{\textsf{random}}\left(S(T_0, \alpha), \ldots, S(T_n, \alpha)\right) & \quad \text{if } T \text{ is a class.} \\
\end{cases}
}\]
In the above, \(R_{\textsf{num}}(x, y)\) selects a value at random from the range \([x, y)\), while \(R_{\textsf{char}}\) produces a character obtained at random.  Similarly to \(I\), for a given method under test \(M\) we define \(S(M)\) as \(S\) applied to each of \(M\)'s parameter types.  We note that \sri with \(\alpha=0\) is in fact equivalent to \ivi, and thus \ivi can be considered a special case of \sri. We note that \thesystem can be configured with different random number distributions to alter the behavior of \(R\).

\subsection{Witness Synthesis}
\label{sec:validation_overview}

Test cases exhibiting abnormal resource consumption are forwarded from the micro-fuzzing phase of the testing procedure to the second phase: witness synthesis and validation.  The rationale behind this phase is to reproduce the behavior during fuzzing in a realistic execution environment using a real JVM in order to avoid false positives introduced due to the measurement instrumentation.

In principle, one could simply interpret any execution that exceeds the configured timeout as evidence of a vulnerability.  In practice, this is an insufficient criterion since the method under test could simply be blocked on I/O, sleeping, or performing some other benign activity.  An additional consideration is that because the \eyevm operates in interpreted mode during the first micro-fuzzing stage (see Section~\ref{sec:ufuzz_impl}), a test case that exceeds the timeout in the first phase might not do so during validation when JIT is enabled.

Therefore, validation of suspected vulnerabilities in a realistic environment is necessary.  To that end, given an abnormal method invocation \(M(v_0, \ldots, v_n)\), a self-contained Java program is synthesized that invokes \(M\) by using a combination of the Reflection API and the Google \texttt{GSON} library.  The program is packaged with any necessary library dependencies and is then executed in a standard JVM with JIT enabled.  Instead of using JVM instrumentation, the wall clock execution time of the entire program is measured.  If the execution was both CPU-bound as measured by the operating system and the elapsed wall clock time exceeds a configured timeout, the synthesized program is considered a witness for a legitimate AC vulnerability and recorded in serialized form in a database.  The resulting corpus of AC vulnerability witnesses are reported to a human analyst for manual examination.

Recall \thesystem takes compiled whole programs and libraries as input. Therefore, the witnesses contained in its final output corpus do not point out the root cause of any vulnerabilities in a program's source code. However, the \eyevm can trace the execution of any Java program running on it (see Section~\ref{subsec:generating_traces}). Given a witness of an AC vulnerability, we can trace its execution in the \eyevm in order to gain insight into the underlying causes of the problem in the program or library. In Section~\ref{sec:evaluation}, we use this technique to discover the root cause for several AC bugs detected by \thesystem.

\section{Implementation}
\label{sec:implementation}

In this section, we describe our prototype implementation of \thesystem and discuss the relevant design decisions. Our prototype implementation consists of \hotfuzzsloc lines of Java code, \eyevmsloc lines of C++ code in the JVM, and \synthsloc lines of Python code for validating AC witnesses detected by micro-fuzzing.

\subsection{\eyevm}
\label{sec:instrumented_jvm}

The OpenJDK includes the HotSpot VM, an implementation of the Java Virtual Machine (JVM), and the libraries and toolchain that support the development and execution of Java programs. The \eyevm is a fork of the OpenJDK that includes a modified HotSpot VM for recording resource measurements. By modifying the HotSpot VM directly, our micro-fuzzing procedure is compatible with any program or library that runs on the OpenJDK.

The \eyevm exposes its resource usage measurement capabilities to analysis tools using the Java Native Interface (JNI) framework.  In particular, a fuzzer running on the \eyevm can obtain the execution time of a given method under test by invoking the \texttt{getRuntime()} method which we added to the existing \texttt{Executable} class in the OpenJDK. The \texttt{Executable} class  allows \fuzzer to obtain a Java object that represents the method under test and access analysis data through our API. This API includes three methods to control and record our analysis: \texttt{setMethodUnderTest}, \texttt{clearAnalysis}, and \texttt{getRuntime}.

We chose to instrument the JVM directly because it allows us to analyze programs without altering them through bytecode instrumentation. This enables us to micro-fuzz a library without modifying it in any way.  It also limits the amount of overhead introduced by recording resource measurements.

The \eyevm can operate in two distinct modes to support our resource consumption analysis: \emph{measurement}, described in Section \ref{subsec:measuring_resource}, and \emph{tracing}, described in Section~\ref{subsec:generating_traces}. In measurement mode, the \eyevm records program execution time with method-level granularity, while tracing mode records method-level execution traces of programs running on the \eyevm. \thesystem utilizes the measurement mode to record the method under test's execution time during micro-fuzzing, while the tracing mode allows for manual analysis of the suspected AC vulnerabilities produced by \thesystem.

\subsubsection{\eyevm Measurement Mode}
\label{subsec:measuring_resource}

Commodity JVMs do not provide a convenient mechanism for recording method execution times.  Prior work has made use of bytecode rewriting~\cite{kuleshov2007using} for this purpose.  However, this approach requires modifying the test artifact, and produced non-trivial measurement perturbation in our testing.  Alternatively, an external interface such as the Serviceability Agent~\cite{RussellB01} or JVM Tool Interface~\cite{jvmti} could be used, but these approaches introduce costly overhead due to context switching every time the JVM invokes a method.  Therefore, we chose to collect resource measurements by instrumenting the HotSpot VM directly.

The HotSpot VM interprets Java programs represented in a bytecode instruction set documented by the JVM Specification~\cite{Java8Spec}.
During start up, the HotSpot VM generates a Template Table and allocates a slot in this table for every instruction given in the JVM instruction set.
Each slot contains a buffer of instructions in the host machine's instruction set architecture that interprets the slot's bytecode.
The Template Interpreter inside the HotSpot VM interprets Java programs by fetching the Java instruction given at the Bytecode Pointer (BCP), finding the instruction's slot in the Template Interpreter's table, and jumping to that address in memory.
The HotSpot VM interprets whole Java programs by performing this fetch, decode, execute procedure starting from the program's entrypoint which is given by a method called \texttt{main} in one of the program's classes.
During execution the Template Interpreter also heavily relies on functionality provided by HotSpot's C++ runtime.

The HotSpot source code contains an Assembler API that allows JVM developers to author C++ methods that, when executed, generate the native executable code required for each slot in the Template Interpreter.
This allows a developer to implement the functionality for a given bytecode instruction, such as \texttt{iadd}, by writing a C++ method \(m\).
When the HotSpot VM starts up, it invokes \(m\), and \(m\) emits as output native code in the host machine's instruction set architecture that interprets the \texttt{iadd} bytecode.
HotSpot saves this native code to the appropriate slot so it can use it later to interpret \texttt{iadd} bytecode instructions.
The API available to developers who author these methods naturally resembles the host's instruction set architecture.
One can think of this Assembler API as a C++ library that resembles an assembler like GNU \texttt{as}.
For example, if the two arguments to an \texttt{iadd} instruction reside in memory, a developer can call methods on this API to load the values into registers, add them together, and store the result on the JVM's operand stack.
We use this API to emit code that efficiently records methods' resource utilization for our analysis.

We instrument the JVM interpreter by augmenting relevant slots in the Template Interpreter using the same API that the JVM developers use to define the interpreter.
To measure execution time, we modify method entry and exit to store the method's elapsed time, measured by the \texttt{RDTSC} Model-Specific Register (MSR) available on the \texttt{x86} architecture, into thread-local data structures that analysis tools can query after a method returns.
We limit our current implementation to the \texttt{x86-64} platform, but this technique can be applied to any architecture supported by the HotSpot VM.
In addition, we could modify the Template Interpreter further to record additional resources, such as memory or disk consumption.

Unfortunately, instrumenting the JVM such that \emph{every} method invocation and return records that method's execution time introduces significant overhead.
That is, analyzing a single method also results in recording measurements for every method it invokes in turn.
This is both unnecessary and adds noise to the results due to both the need to perform an additional measurement for each method as well as the adverse effects on the cache due to the presence of the corresponding measurement field.
Thus, our implementation avoids this overhead by restricting instrumentation to a single \methodundertest that \fuzzer can change on demand.

In particular, \fuzzer stores the current \methodundertest inside thread-local data.
During method entry and exit, the interpreter compares the current method to the thread's \methodundertest.
If these differ, the interpreter simply jumps over our instrumentation code.
Therefore, any method call outside our analysis incurs at most one comparison and a short jump.

Every time the interpreter invokes a method, our instrumentation stores the latest value of \rdtsc into an attribute $T_{start}$ in the calling thread and increments a depth counter $T_{depth}$. If the same method enters again in a recursive call, we increment $T_{depth}$.  If the \methodundertest calls another method, it simply skips over our analysis code.  Each time the \methodundertest returns, we decrement $T_{depth}$. If $T_{depth}$ is equal to zero, the \eyevm  invokes \rdtsc and the computed difference between the current value and $T_{start}$ is stored inside the calling thread.  Observe that the measured execution time for the \methodundertest consequently includes its own execution time and the execution time of all the methods it invokes.  This result is stored inside the \methodundertest's internal JVM data structure located in its class's constant pool.
The Assembler API available in the JVM sources supports all the functionality needed to implement these measurements, including computing the offsets of C++ attributes, manipulating machine registers, and storing values to memory.

Every time the JVM invokes a method, the Template Interpreter sets up a new stack frame for the method which the interpreter removes after the method returns.
The code that implements this logic is defined using the same Assembler API that implements each JVM bytecode instruction.
To record our resource measurements, we insert relevant code snippets into the Template Interpreter that run every time the \eyevm adds or removes a stack frame.

The \texttt{java} executable that runs every Java program begins by loading the HotSpot VM as a shared library into its process address space in order to run the JVM. Thus, \eyevm can export arbitrary symbols to expose a JNI interface to analysis tools implemented in Java. Currently, the \eyevm defines functions that allow a process to configure the \methodundertest, poll the method's most recent execution time, and to clear the method's stored execution time. The \eyevm then simply uses the JNI to bind the methods we added to the \texttt{Executable} Java class to the native \eyevm functions that support our analysis.

\subsubsection{\eyevm Tracing Mode}
\label{subsec:generating_traces}

In addition to measuring method execution times, \eyevm allows an analyst to
trace the execution of Java programs with method-level granularity. Tracing
provides valuable insight into programs under test and is used herein to
evaluate \thesystem's ability to detect AC vulnerabilities (see
Section~\ref{sec:evaluation}). Each event given in a trace represents either a
method invocation or return. Invocation events carry all parameters passed to
the method as input.

In principle, traces could be generated either by instrumenting the bytecode of the program under test, or through an external tool interface like the JVMTI. As both of these approaches introduce significant overhead, we (as for measurement mode) opt instead for JVM-based instrumentation.  That is, modifying the JVM directly to trace program execution does not require any modification of the program under analysis and only requires knowledge of internal JVM data structures.

\eyevm's tracing mode is implemented by instrumenting the bytecode interpreter generated at run-time by the HotSpot VM. Recall that the JVM executes bytecode within a generated Template Interpreter in the host machine's instruction set architecture. In order to generate program traces that record all methods invoked by the program under test, stubs are added to the locations in the Template Interpreter that invoke and return from methods.  We note that these are the same locations that are instrumented to implement measurement mode.

However, while performance overhead is an important factor, program execution tracing can nevertheless be effectively implemented in the C++ run-time portion of the JVM as opposed to generating inline assembly as in the measurement case.  Then, during interpreter generation, all that is added to the generated code are invocations of the C++ tracing functions.

To trace a program under test, we define a trace recording point as when the program either invokes a method or returns from one.
When a \methodundertest reaches a trace recording point the JVM is executing in the generated Template Interpreter represented in x86-64 assembly. Directly calling a C++ function in this state will lead to a JVM crash, as the machine layout of the bytecode interpreter differs from the Application Binary Interface (ABI) expected by the C++ run-time. Fortunately, the JVM provides a convenient mechanism to call methods defined in the C++ run-time using the \texttt{call\_VM} method available in the Assembler API. The \texttt{call\_VM} method requires that parameters passed to the C++ function are contained within general purpose registers. This facility is used to pass a pointer to the object that represents the method we wish to trace, a value that denotes whether the event represents an invocation or return, and a pointer to the parameters passed to the method under test.  All of this information is accessible from the current interpreter frame when tracing an event. The JVM maintains an Operand Stack that holds inputs to methods and bytecode instructions. Internally, a special variable called the Top of the Stack State (ToSState) allows the JVM to check where the top of the Operand Stack is located. Before calling our C++ stub to trace an event, we push the current ToSState onto the machine stack. Next, we call our C++ tracing function. After the tracing function returns, we pop the ToSState off the machine stack and restore it to its original value.

The trace event stub itself collects the name of every invoked method or constructor, and its parameters. The name of the method is obtained from the method object the JVM passes to the stub.  The parameters passed to the method under test are collected by accessing the stub parameters in similar fashion. The JVM's \texttt{SignatureIterator} class allows the tracing function to iterate over the parameter types specified in the method under test's signature, and, therefore, ensures that tracing records the correct parameter types.  For each parameter passed to a method, both its type and value are saved.  Values of primitive types are represented as literals, whereas objects are represented by their internal ID in the JVM. Within the trace file, one can find the origin of a given ID from the object's constructor in the trace.  All of this information is streamed to a trace file one event at a time.

\subsection{\fuzzer}
\label{sec:ufuzz_impl}

Micro-fuzzing is implemented using a message broker and a collection of \fuzzer
instances.  Each \fuzzer instance runs inside the \eyevm in measurement mode,
consumes methods as jobs from a queue, and micro-fuzzes each method within its
own process. Over time, micro-fuzzing methods in the same process might
introduce side-effects that prevent future jobs from succeeding. For example, a
method that starts an applet could restrict the JVM's security policy and
prevent \fuzzer from performing benign operations required to fuzz future
methods.  This occurs because once a running VM restricts its security policy,
it cannot be loosened.  To prevent this and similar issues from affecting
future micro-fuzzing jobs, we add the following probe to every \fuzzer
instance. Prior to fuzzing each method received from the job queue, \fuzzer
probes the environment to ensure basic operations are allowed. If this probing
results in a security exception, the \fuzzer process is killed and a new one is
spawned in its place. Traditional fuzzers avoid these problems by forking
before each test case so it can run in a fresh state. For a simple Java program
that loops indefinitely, the JVM runs 16 operating system threads. Constantly
forking such a heavily multi-threaded environment on every test case introduces
unnecessary complexity and very quickly destabilizes our experiments.

We configure each \fuzzer instance in the following way to prevent
non-determinism present in the JVM from introducing unnecessary noise into our
experiments.  Every \fuzzer instance runs within the \eyevm in interpreted mode
in order to maintain consistent run-time measurements for methods under test.
If \fuzzer runs with JIT enabled, our measurement instrumentation no longer
profiles the method under test, but rather the JVM's response to fuzzing the
method under test. A JVM with JIT enabled responds by compiling the bytecode
that implements the method under test into equivalent native code in the host
machine's instruction set architecture and executes it in a separate code
cache. This causes the method under test's runtime to change dramatically
during micro-fuzzing and would skew our results. For this reason, we run
\fuzzer in the \eyevm in interpreted mode to ensure consistent measurements.

Upon receiving a method under test, \fuzzer queries the CPUs available by
obtaining the process's CPU affinity mask with \texttt{sched\_getaffinity}.
\fuzzer then calls \texttt{sched\_setaffinity} to pin the thread running the
method under test to the lowest CPU given in the affinity mask. This confines
the method under test to a single CPU for the duration of micro-fuzzing and also
requires that every \fuzzer instance have access to at least two CPUs, one
for the method under test, and the remainder for the JVM's own threads.

Each time \fuzzer successfully invokes the method under test, it submits a test
case for storage in the results database.  Every test case generated by \fuzzer
consists of the input given to the method under test and the number of clock
cycles it consumes when invoked on the input. \fuzzer interprets exceptions as
a signal that an input is malformed, and therefore all such test cases are
discarded. Ignoring input that causes the \methodundertest to throw an
exception restricts \fuzzer's search space to that of valid inputs while it
attempts to maximize resource consumption. In a different context, these test
cases could be considered a potential attack vector for triggering DoS, but not
due to an AC vulnerability.

\section{Evaluation}
\label{sec:evaluation}

In this section, we describe an evaluation of our prototype implementation of
\thesystem. This evaluation focuses on the testing procedure's efficiency in
finding AC vulnerabilities in Java libraries, and additionally considers the
effect of seed input instantiation strategy on micro-fuzzing efficiency.  In
particular, we define the performance of micro-fuzzing as the number of AC
vulnerabilities detected in a test artifact over time, and consider one
strategy to outperform another if the strategy detects more AC vulnerabilities
given the same time budget. In accordance with recently proposed guidelines
for evaluating new fuzz testing techniques~\cite{KleesRCW018}, we evaluate our proposed
seed selection strategy (\sri) by comparing the performance of micro-fuzzing with
\sri-based seeds to micro-fuzzing with ``empty seed values'' (\ivi-based seeds).

To the best of our knowledge, including existing fuzzers in our evaluation
that adapt AFL~\cite{guidovranken} and libFuzzer~\cite{isstac} to Java programs
to detect AC vulnerabilities would require significant engineering effort both
in terms of instrumentation and designing test harnesses around our artifacts.
Furthermore, the results those tools have achieved on real world code bases
like the JRE or Java libraries appear limited to individual methods and find
bugs that crash the \methodundertest with a thrown exception, or individual
challenges from the STAC program. For these reasons, we exclude those tools
from our evaluation.

We evaluate \thesystem over the Java Runtime Environment (JRE), all challenge
programs developed by red teams in the \stac program, and the 100 most popular
libraries available on the Maven repository.
This set of evaluation artifacts presents the opportunity to detect previously
unknown vulnerabilities in real-world software as well as to validate
\thesystem on programs for which we have ground truth for AC vulnerabilities.

For the real-world software evaluation, we selected the JRE as it provides basic functionality utilized by every Java program. Given Java's widespread deployment across domains that range from embedded devices to high performance servers, any unknown AC vulnerabilities in the JRE present significant security concerns to programs that utilize those methods.  For this reason, we evaluate \thesystem over all methods in the JRE in order to measure its ability to detect unknown AC vulnerabilities in production software.  Specifically, we consider JRE \jreversion from Java~8 as a library under test in our evaluation.

In addition to the JRE, Java programs frequently rely on libraries available through the popular Maven Repository, which as of 2019 hosts 15.1 million artifacts.
While this repository provides a convenient way to download an application's library dependencies, it also introduces any unknown vulnerabilities hiding within them into an application.
In order to understand how vulnerable Maven's libraries are to AC attacks, we evaluate \thesystem over the repository's 100 most popular libraries.
A library's popularity on Maven is defined by the number of artifacts that include it as a dependency.
For every Maven library we consider in our evaluation, we micro-fuzz every method contained in the library, and exclude the methods contained in its dependencies.

\paragraph{Findings Summary}
In conducting our evaluation, \thesystem detected previously unknown AC vulnerabilities in the JRE, Maven Libraries, and discovered both intended and unintended AC vulnerabilities in STAC program challenges. Section~\ref{subsec:setup} documents the experimental setup used to obtain these findings. Section~\ref{subsec:results} summarizes how the seed input generation strategy impacts micro-fuzzing performance, and provides several detailed case studies of micro-fuzzing results for the JRE, STAC challenges, and Maven libraries.

\subsection{Experimental Set Up}
\label{subsec:setup}

\begin{table*}[t]

\caption{The parameters given to every \fuzzer instance. Multiple timeouts prevent \thesystem from stopping because of individual methods that may be too difficult to micro-fuzz efficiently.}
\begin{center}
\begin{tabular}{llr}
   \toprule
   Parameter & Definition & Value \\
   \midrule
   $\alpha$ & The maximum value \sri will assign to a primitive type when constructing an object  & 256 \\
   $\psi$ & The maximum amount of time to create the initial population & 5s \\
   $\lambda$ & The time that may elapse betwen measuring the fitness of two method inputs & 5s \\
   $\omega$ & The amount of time required for a method to run in order to generate an AC witness & 10s \\
   $\gamma$ & The wall clock time limit for the GA to evaluate the method under test. & 60s \\
   $\pi$ & The size of the initial population & 100 \\
   $\chi$ & The probability two parents produce offsprint in a given generation & 0.5 \\
   $\tau$ & The probability an individual mutates in a generation & 0.01 \\
   $\epsilon$ & The percent of the most fit individuals that carry on to the next generation & 0.5 \\
   $\nu$ & The number of generations to run the GA & 100 \\
   $\sigma$ & The amount of time required for an AC witness to run outside the analysis framework in JIT mode in order to confirm it & 5s \\
   \bottomrule
\end{tabular}

\end{center}
\label{tbl:evaluation_parameters}
\end{table*}

We implement \thesystem as a distributed system running within an on-premise
Kubernetes cluster. The cluster consists of 64~CPUs and 256~GB of RAM across 6
Dell PowerEdge R720 Servers with Intel Xeon 2.4~GHz Processors. To micro-fuzz a
given library under test, we deploy a set of \fuzzer instances onto the cluster
that consume individual methods from a message broker.

For each individual method under test, each \fuzzer instance creates an initial
population of inputs to the method, and runs a genetic algorithm that searches
for inputs that cause the method under test to consume the most execution time.
In addition to the method under test, every job submitted to \thesystem
requires configuration parameters given in Table~\ref{tbl:evaluation_parameters}.

Recall that \thesystem makes no assumptions about the code under test, and
therefore it is critical to configure timeouts at each step of this process in
order for the whole system to complete for all methods in a library under test.
For this reason, the parameters \(\left(\psi, \lambda, \omega, \gamma\right)\)
configured various timeouts that ensured \thesystem ran end to end within a
manageable time frame. This is important in order to prevent problems caused by
fuzzing individual methods or calling specific constructors from halting the
entire micro-fuzzing pipeline. We determined the values for these parameters
empirically, and only added each parameter after we observed a need for each
one in our evaluation.
The parameters \(\left(\pi, \chi, \tau, \epsilon, \nu\right)\) configured the
genetic algorithm (GA) within \thesystem. In our evaluation, we assigned these
parameters the initial values recommended for genetic algorithm
experiments~\cite{EibenS15}.
Finally, \(\sigma\) configured the timeout used in the witness validation
stage.  Observe that we configured \(\sigma\), the time required to confirm a
witness as an AC vulnerability, to be half of \(\omega\), the time needed to
detect a witness.  Our intuition behind this choice is that a given test case
will run much faster with JIT enabled than in our interpreted analysis
environment, and hence the runtime required to confirm a witness is lower than
the time required to detect it.  We use the same parameters for every method
under test and do not tune these parameters for specific methods. We argue that
this provides a generic approach to detecting AC vulnerabilities in arbitrary
Java methods.

To micro-fuzz each library under test, we created a pair of fuzzing jobs with
identical parameters for each method contained in the library with the
exception of the \(\alpha\) parameter. Each pair consisted of one job that used
the \ivi seed input generation strategy, and the other used the \sri strategy
with the \(\alpha\) parameter which bounds the values used when constructing
the seed inputs for micro-fuzzing.
The libraries under test that we consider for our evaluation are all
\stacchallenges engagement articles given in the STAC program, and every public
method contained in a public class found in the JRE and the 100 most popular
libraries available on the Maven repository.  For the latter, we consider these
public library classes and methods as the interface the library reveals to
programs that utilize it. Therefore, this provides an ideal attack surface for
us to micro-fuzz for potential AC vulnerabilities.

\paragraph{Seed Input Strategy}
\label{sec:seed_population}

Given the definition of \sri presented in Section~\ref{sec:sri}, we use the
following procedure to construct the initial seed inputs for every method under
test \(M\) configured to use \sri in our evaluation. Given the parameter
\(\alpha\), \fuzzer instantiates a population
of seed inputs of size \(\pi\) for \(M\) as follows. Let \(\mathcal{N}(\mu, \sigma)\)
be a normal random number distribution with mean \(\mu\) and standard deviation
\(\sigma\). For every primitive type required to instantiate a given class,
\sri obtains a value \(X \leftarrow \mathcal{N}(0, \nicefrac{\alpha}{3})\).
This allows micro-fuzzing to favor small values centered around 0 that exceed
the chosen parameter \(\alpha\) with small probability. To be precise,
configuring \(\alpha\) to be three times the size of the standard deviation
\(\sigma\) of our random number distribution \(\mathcal{N}\) makes \(Pr(|X| >
\alpha) < 0.135 \%\). This ensures that the primitive values we use to
instantiate objects stay within the range \(\left[-\alpha, \alpha\right]\) with
high probability.

\subsection{Experimental Results}
\label{subsec:results}

\begin{figure*}[t!]
  \centering
  \begin{subfigure}[t]{0.5\textwidth}
    \includegraphics[angle=270,width=\textwidth]{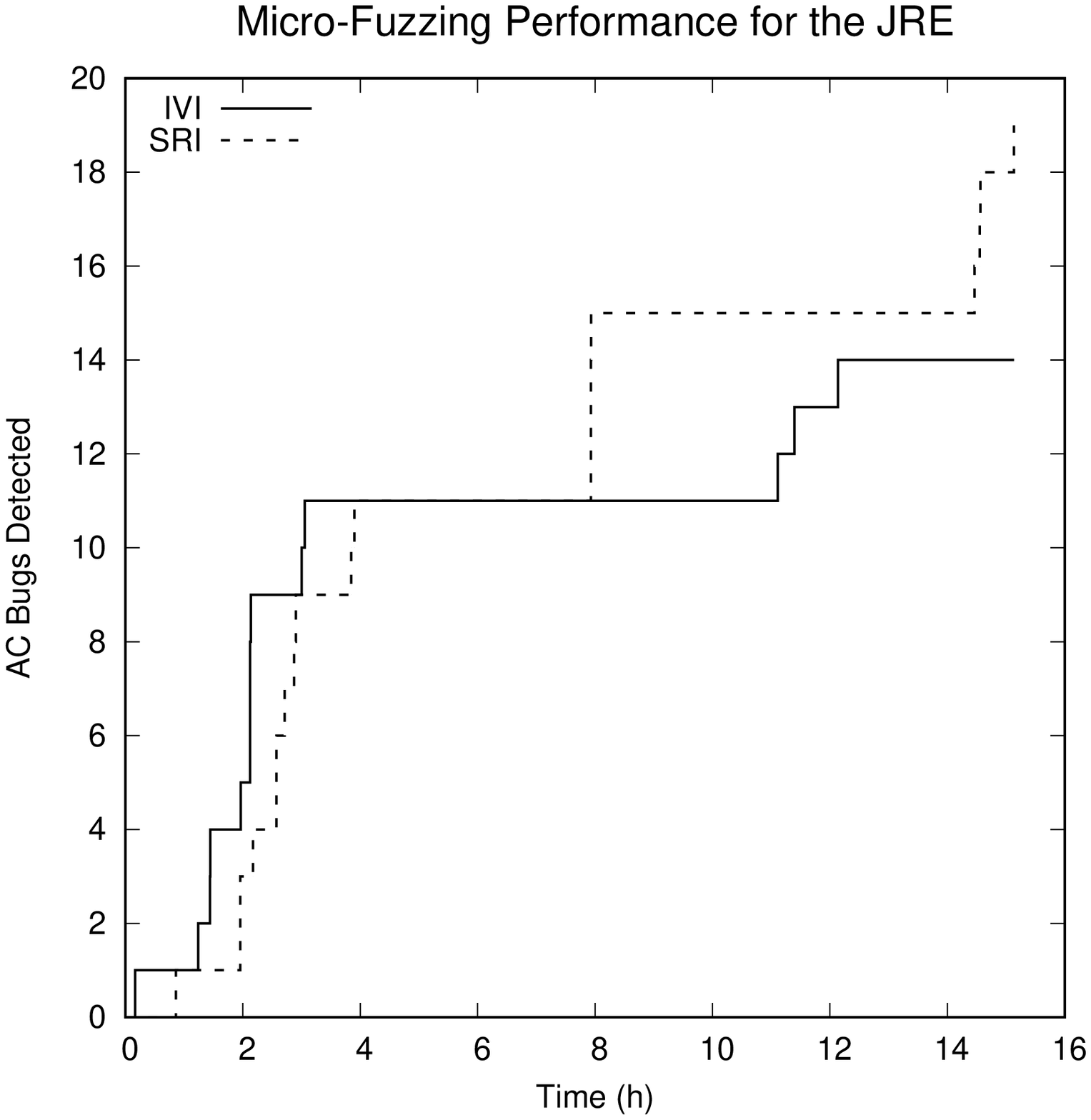}
    \caption{}
    \label{perf:jre}
  \end{subfigure}%
  \hfill
  \begin{subfigure}[t]{0.5\textwidth}
    \includegraphics[angle=270,width=\textwidth]{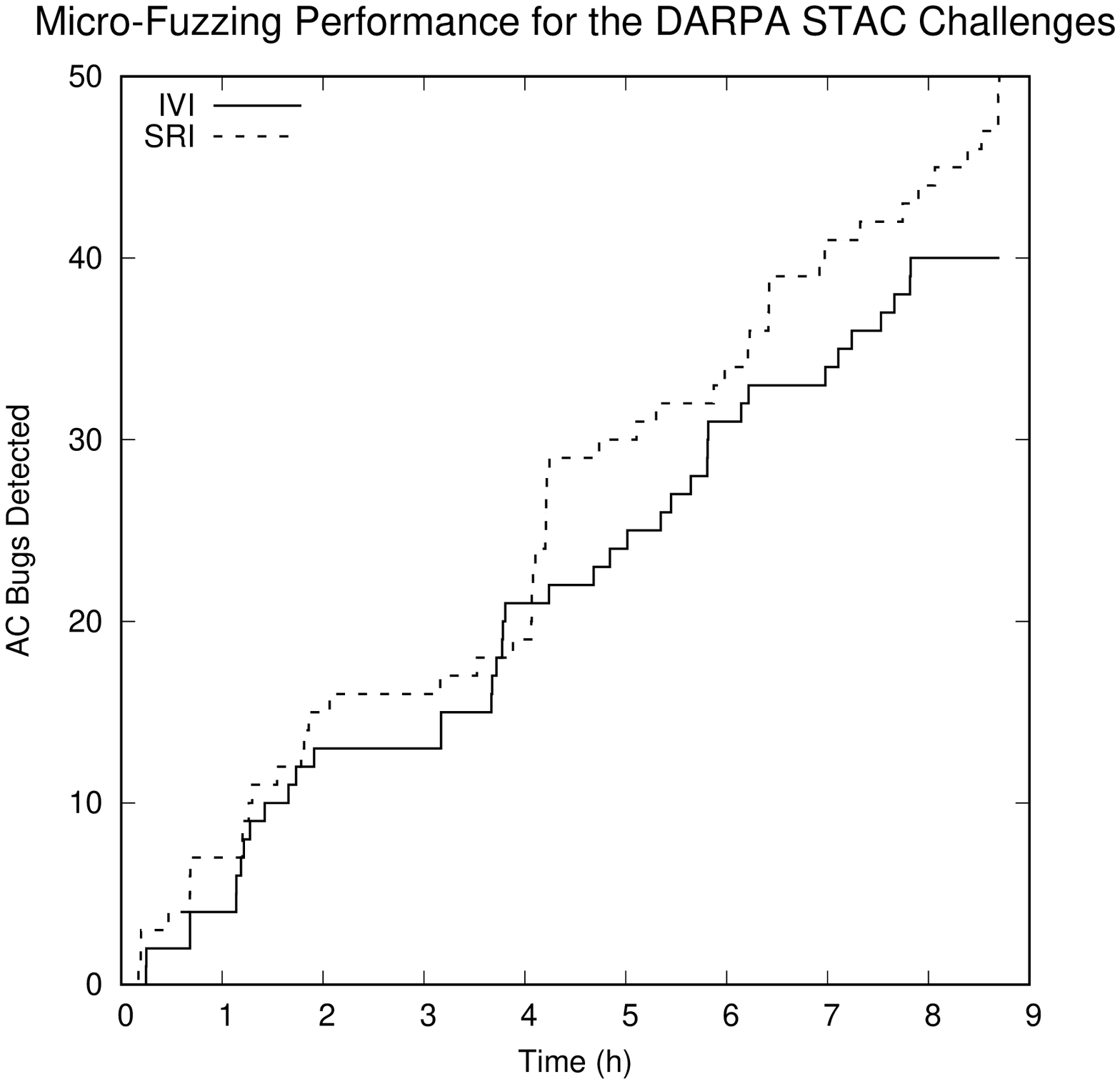}
    \caption{}
    \label{perf:darpa}
  \end{subfigure}
  \hfill
  \begin{subfigure}[t]{0.5\textwidth}
    \includegraphics[angle=270,width=\textwidth]{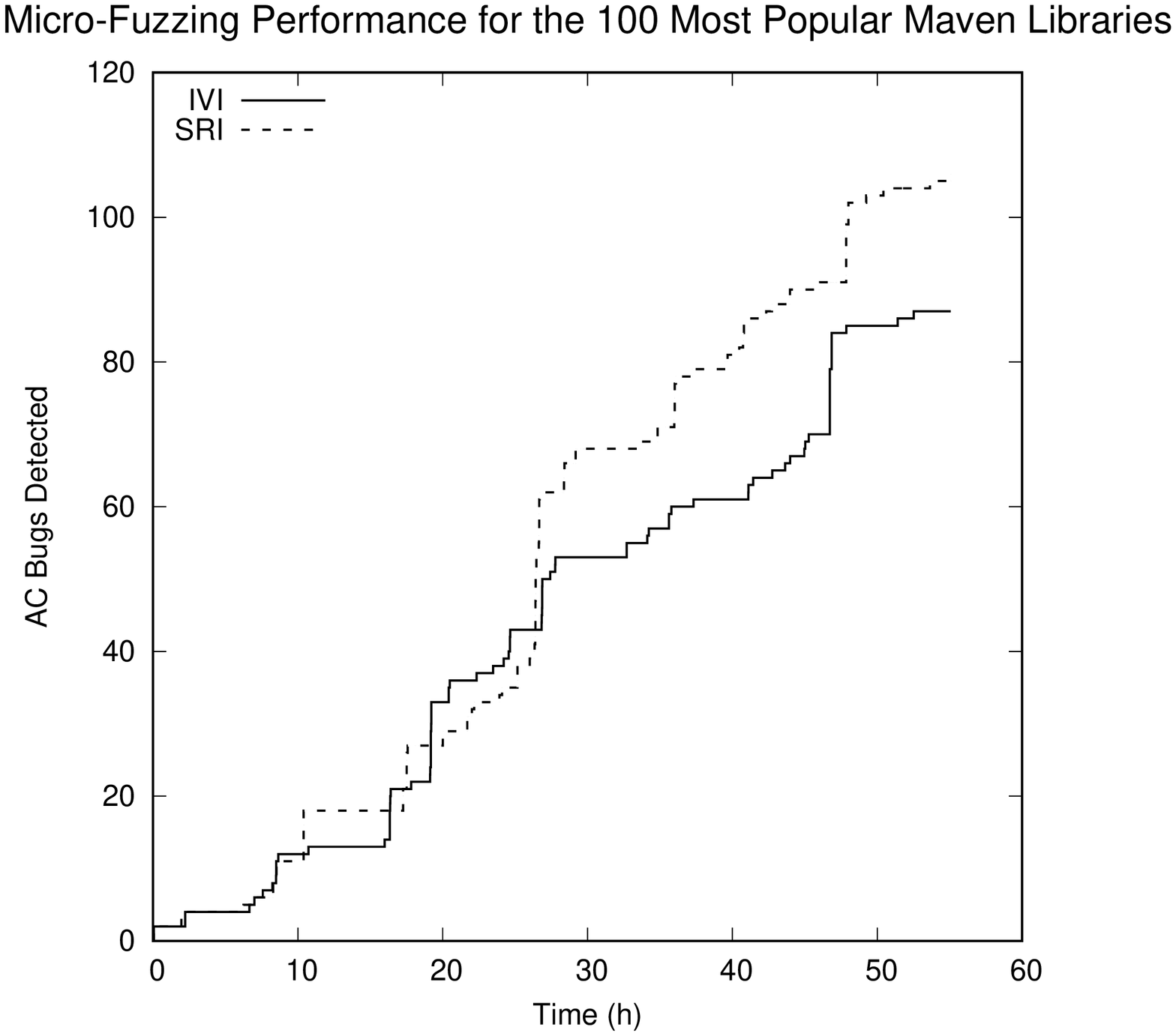}
    \caption{}
    \label{perf:maven}
  \end{subfigure}
\caption{Measuring \thesystem's performance while micro-fuzzing the Java
         Runtime Environment (JRE) (Figure~\ref{perf:jre}), all of the challenges given in the \stac program (Figure~\ref{perf:darpa}), and the 100 Most Popular Libraries found in Maven~\cite{maven} (Figure~\ref{perf:maven}) using both \ivi and \sri-derived seed inputs. Observe that micro-fuzzing with \sri-derived seed inputs outperforms \ivi-derived seed inputs for every artifact contained in our evaluation.}
    \label{fig:hotfuzzjreperformance}
\end{figure*}

In our evaluation, \thesystem detected \zerodaynum previously unknown AC
vulnerabilities in the Java~8 JRE, detects both intended and unintended
vulnerabilities in challenges from the STAC program, and detects
\mavenzerodaynum AC vulnerabilities in \vulnmavenlibraries libraries from the
100 most popular libraries found on Maven. Table~\ref{tbl:seed_evaluation}
breaks down both the total wall-clock time \thesystem spent micro-fuzzing the
JRE, STAC engagement challenges, and Maven libraries for each seed selection
strategy and reports micro-fuzzing's throughput measured by the average number
of test cases produced per hour. We define a test case to be a single input
generated by \thesystem for a method under test.  Overall, micro-fuzzing with
\sri-derived seed inputs required more time to micro-fuzz the artifacts in our
evaluation, but also produced more test cases overall.

\subsubsection{Impact of Seed Input Generation Strategy}

\newcommand{\onlyemptybugs}{26 \xspace}
\newcommand{\sribugs}{23 \xspace}
\newcommand{\srinonbugs}{3 \xspace}

\begin{table*}[t]
\caption{A comparison of fuzzing outcomes when evaluating \thesystem on Java libraries using IVI and SRI seed inputs.}
\begin{center}
\begin{tabular}{lcccccccccccccc}
  \toprule
  \multirow{1}{*}{Library} & \multicolumn{1}{c}{Size} & \multicolumn{3}{c}{AC Witnesses Detected} & \multicolumn{3}{c}{AC Witnesses Confirmed} & \multicolumn{3}{c}{Methods Covered} & \multicolumn{2}{c}{Fuzzing Time (hrs)} & \multicolumn{2}{c}{Throughput (tests/hr)} \\
  \cmidrule(lr){2-2} \cmidrule(lr){3-5} \cmidrule(lr){6-8} \cmidrule(lr){9-11} \cmidrule(lr){12-13} \cmidrule(lr){14-15}
   & \multicolumn{1}{l}{No. Methods} & \multicolumn{1}{c}{Both} & \multicolumn{1}{c}{IVI} & \multicolumn{1}{c}{SRI} & \multicolumn{1}{c}{Both} & \multicolumn{1}{c}{IVI} & \multicolumn{1}{c}{SRI} & \multicolumn{1}{c}{Both} & \multicolumn{1}{c}{IVI} & \multicolumn{1}{c}{SRI} & \multicolumn{1}{c}{IVI} & \multicolumn{1}{c}{SRI} & \multicolumn{1}{c}{IVI} & \multicolumn{1}{c}{SRI}\\
\midrule
\library{JRE} & 91632 & 6 & 8 & 13 & 5 & 8 & 13 & 23818 & 2780 & 1439 & 14.3 & 19.5 & 4389675 & 3092866\\
\library{DARPA STAC} & 67494 & 34 & 6 & 15 & 5 & 0 & 0 & 8064 & 847 & 1162 & 7.8 & 8.7 & 3608741 & 3172502\\
\library{Top 100 Maven Libraries} & 239777 & 46 & 38 & 56 & 43 & 35 & 54 & 66987 & 2622 & 1770 & 52.9 & 55.5 & 5906687 & 5591106\\
\bottomrule
\end{tabular}
\end{center}
\label{tbl:seed_evaluation}
\end{table*}

Table~\ref{tbl:seed_evaluation} presents the results of micro-fuzzing the JRE,
all the challenges contained in the DARPA STAC program, and the 100 most popular
libraries available on Maven using both \ivi and \sri-derived seed inputs.
Overall, micro-fuzzing with both strategies
managed to invoke \jremethodscovered of the methods contained in the JRE,
\stacmethodscovered of the methods given in the STAC program, and
\mavenmethodscovered of the methods found in the 100 most popular Maven
libraries.
As the results indicate, neither seeding strategy is categorically superior to
the other, although \sri does consistently outperform \ivi on each of the
artifacts included in our evaluation. For example, when considering the
results over the JRE, \sri identifies 13 vulnerabilities that \ivi does not,
compared to the 8 vulnerabilities \ivi finds that \sri does not. At the same time,
both strategies are able to find another 5 vulnerabilities.
We observe the same pattern in both the STAC artifacts and Maven libraries we
included in our evaluation, although observe that both strategies were able to
solve the same number of challenges in our experiment over the STAC articles
(see Case Study~\ref{subsec:darpa_eval}).
These results indicate that \sri outperforms \ivi as a seed input strategy, and
overall the two approaches are \emph{complementary}.  Furthermore, their
combined results show that relatively simple micro-fuzzing can expose serious
availability vulnerabilities in widely-used software.

Figure~\ref{fig:hotfuzzjreperformance} visually compares the performance of
micro-fuzzing the JRE, STAC challenges, and the 100 most popular Maven
libraries, using both \ivi and \sri-derived seed inputs. From these results, we
see that \sri-derived seed inputs produce a marginal improvement over \ivi
inputs, as micro-fuzzing with \sri-derived seed inputs detects more AC bugs
than \ivi seed inputs in each of our evaluation artifacts.

\begin{figure*}[t]
    \centering
    \begin{subfigure}{0.46\textwidth}
        \includegraphics[angle=270,width=\columnwidth]{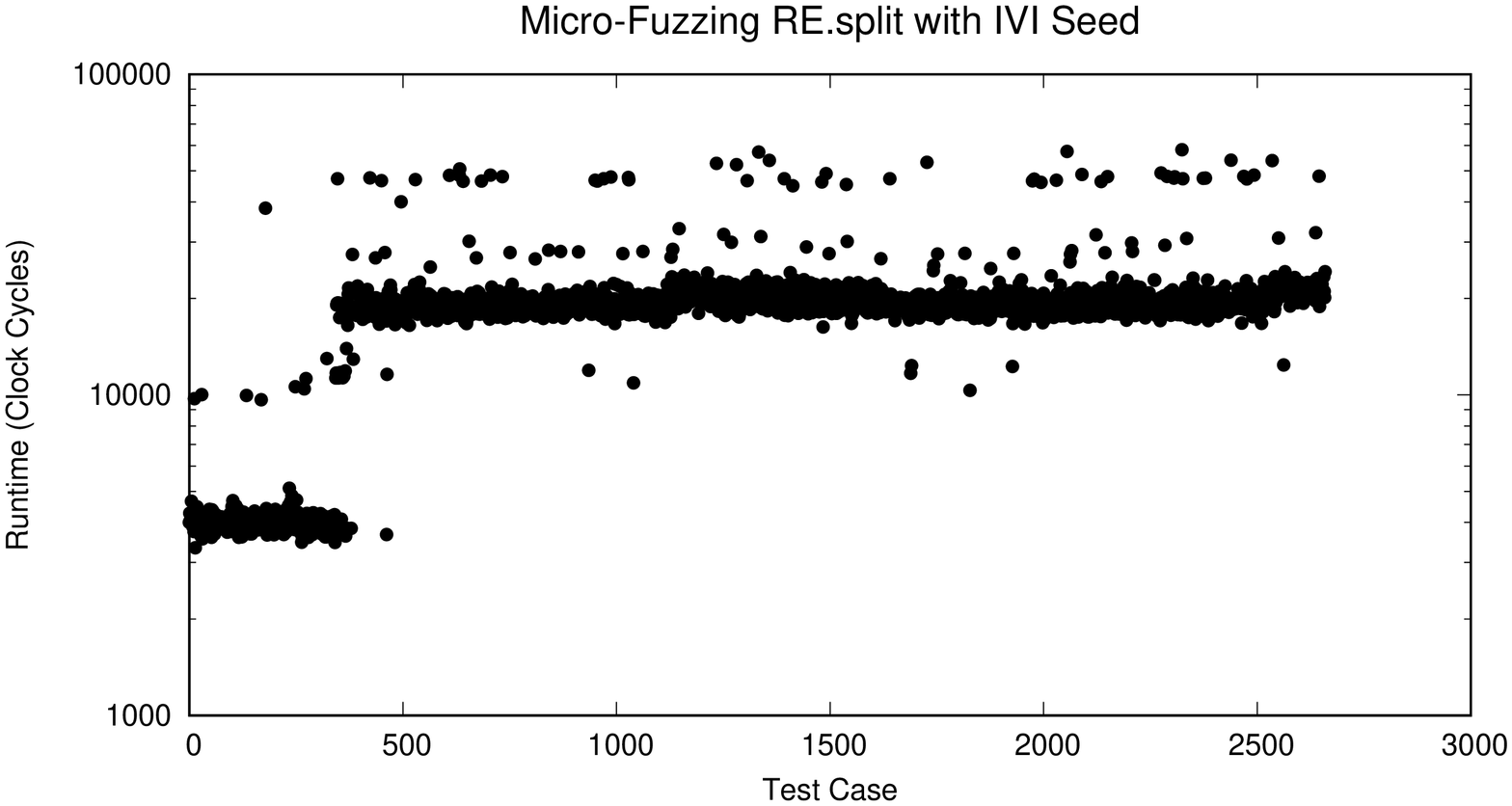}
        \caption{\ivi-derived inputs.}\label{fig:re_empty_seed}
    \end{subfigure}
    \qquad
    \begin{subfigure}{0.46\textwidth}
        \includegraphics[angle=270,width=\columnwidth]{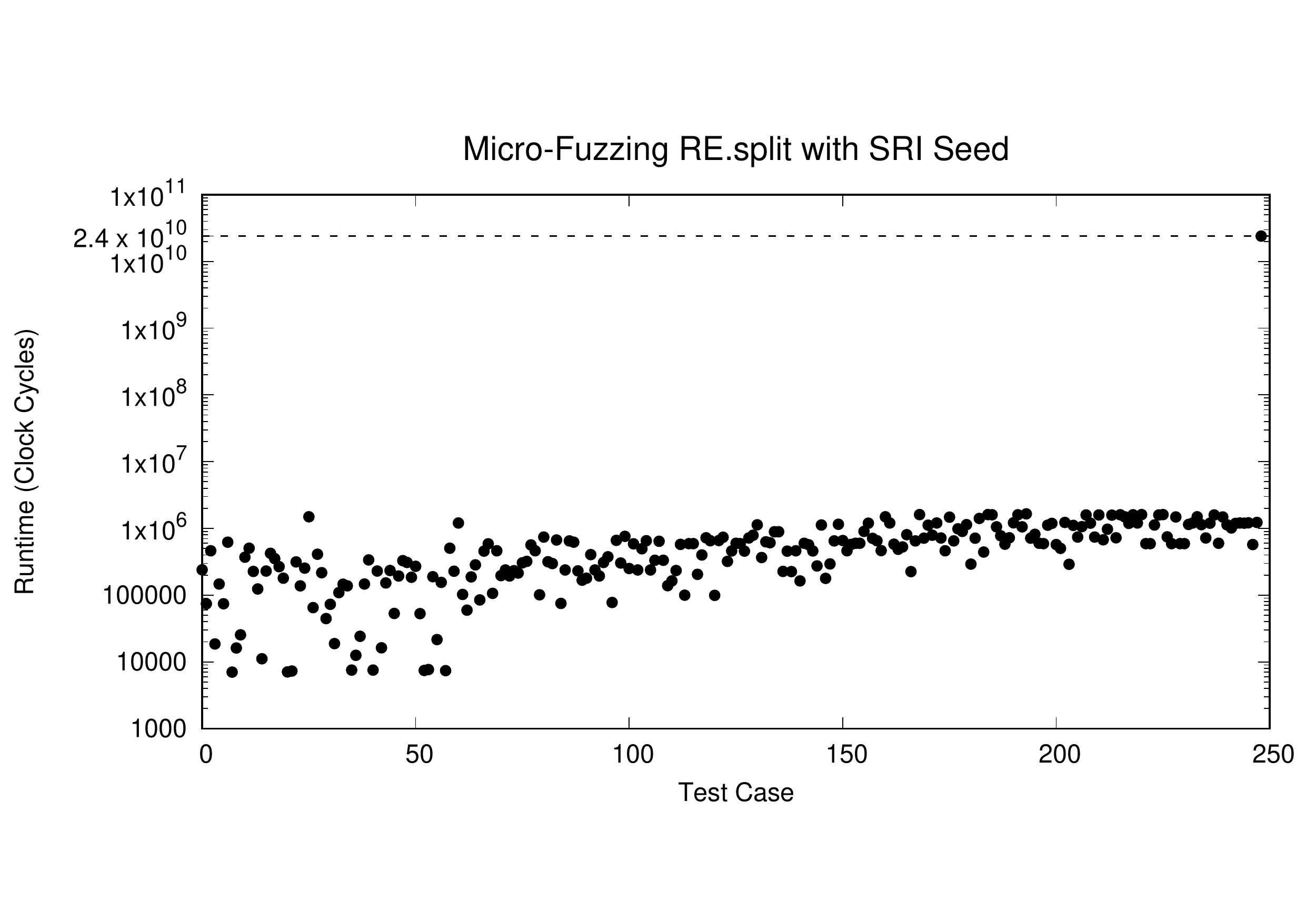}
        \caption{\sri-derived inputs.}\label{fig:re_sri_seed}
    \end{subfigure}
    \caption{Visualization of micro-fuzzing {\tt
    com.sun.org.apache.regexp.internal.RE.split(String input)} with \ivi and \sri-derived inputs.
     Micro-fuzzing with \ivi-derived inputs fails to detect a zero-day AC
     vulnerability in the JRE, while \sri-derived inputs detect the
     vulnerability correctly. Observe that the y-axis for graph~\subref{fig:re_empty_seed} is 5~orders of magnitude smaller than graph~\subref{fig:re_sri_seed}. Graph~\subref{fig:re_sri_seed} shows the test case that triggers the AC vulnerability in the upper-right-hand corner.}
    \label{fig:seedcomparisongroup}
\end{figure*}

Figure~\ref{fig:seedcomparisongroup} provides a visual comparison between
\ivi and \sri-based micro-fuzzing on a method provided by the JRE that
works on regular expressions. According to the documentation, the {\tt
RE.split(String input)} method splits a given string {\tt input} into an array
of strings based on the regular expression boundaries expressed in the compiled
regular expression instance {\tt RE}.
Figure~\ref{fig:re_empty_seed} shows how micro-fuzzing this method using
\ivi-based seeds fails to arrive at a test case that demonstrates the
vulnerability. In contrast, Figure~\ref{fig:re_sri_seed} shows how using
\sri-based seed inputs allows \thesystem to detect the vulnerability.
Additionally, we note that micro-fuzzing with \sri-derived seed inputs requires
fewer test cases than micro-fuzzing with \ivi-based seeds. When we traced the
execution of the exploit found by \thesystem in the \eyevm in tracing mode, we
discovered that the method called the \texttt{StringCharacterIterator.isEnd}
method from the \texttt{com.sun.org.apache.regexp.internal} package with
alternating arguments indefinitely. We observed the PoC produced by \thesystem
run for \resplittime on a Debian system with Intel Xeon E5-2620 CPUs before
stopping it. We have disclosed our PoC to Oracle and await their response.

Our evaluation revealed a second AC vulnerability within the same \texttt{RE}
package called \texttt{subst(String substituteIn, String substitution)} that
substitutes every occurrence of the compiled regular expression in
\texttt{substituteIn} with the string \texttt{substitution}. After tracing the
PoC produced by \thesystem, we observed that it has the same underlying problem
as \texttt{RE.split}. That is, it appears to loop indefinitely checking the result
of \texttt{StringCharacterIterator.isEnd}. We observed the PoC for
\texttt{RE.subst} running for 12 days on the same Debian system on which we
tested the \texttt{RE.split} PoC before stopping it. After reporting the issue
to Oracle, they claimed it is not a security issue since the \texttt{RE.subst}
method is protected and an attacker would have to perform a non-trivial amount
of work to access it. That being said, the test case generated by \thesystem is
only 579 bytes in size and no method in the OpenJDK sources utilizes the
\texttt{RE.subst} method outside of the OpenJDK test suite.  This method
appears to serve no purpose beyond providing a potential attack surface for
DoS.

\subsubsection{Case Study: Detecting AC Vulnerabilities with \ivi-based Inputs}
\label{sec:empty_eval}

Our evaluation revealed the surprising fact that \jreemptybugs methods in the JRE contain AC vulnerabilities exploitable by simply passing empty values as input. Figure~\ref{fig:emptyexploits} shows the \jreemptybugs utilities and APIs that contain AC vulnerabilities that an adversary can exploit. Upon disclosing our findings to Oracle they communicated that five of the six methods (lines 1-29) belong to internal APIs and that no path exists for malicious input to reach them.
They recognized \texttt{DecimalFormat}'s behavior (lines 31-34) as a functional bug that they will fix in an upcoming release.
Oracle's assessment assumes that a malicious user will not influence the input of the public \texttt{DecimalFormat} constructor. Unless programs exclusively pass string constants to the constructor as input, this is a difficult invariant to always enforce.

\subsubsection{Case Study: Arithmetic DoS in Java Math}
\label{sec:jdkmath_eval}

\begin{figure}[]
    \centering
    \begin{lstlisting}[
      language=Java
    ]
clojure=> (inc (BigDecimal. "1e2147483647"))
clojure=> (dec (BigDecimal. "1e2147483647"))

groovy:000> 1e2147483647+1

scala> BigDecimal("1e2147483647")+1

JSONObject js = new JSONObject();
js.put("x", BigDecimal("1e2147483647"));
js.increment("x");
\end{lstlisting}
    \caption{Proof of concept exploits that trigger inefficient arithmetic operations for the \texttt{BigDecimal} class in Clojure, Scala, Groovy, and the org.json library.}
    \label{fig:bigdecimalpoc}
\end{figure}

As a part of our evaluation, HotFuzz detected 5~AC vulnerabilities inside the JRE's Math package.  To the best of our knowledge, no prior CVEs document these vulnerabilities. We developed proof-of-concept exploits for these vulnerabilities and verified them across three different implementations of the JRE from Oracle, IBM, and Google. The vulnerable methods and classes provide abstractions called \texttt{BigDecimal} and \texttt{BigInteger} for performing arbitrary precision arithmetic in Java. Any Java program that performs arithmetic over instances of \texttt{BigDecimal} derived from user input may be vulnerable to AC exploits, provided an attacker can influence the value of the number's exponent when represented in scientific notation.

A manually defined exploit on \texttt{BigDecimal.add} in Oracle's JDK (Versions~9 and~10) can run for over an hour even when Just-in-Time (JIT) compilation is enabled.  On IBM's J9 platform, the exploit ran for \jninetime, as measured by the \texttt{time} utility, before crashing.  When we exploit the vulnerability on the Android 8.0 Runtime (ART), execution can take over 20~hours before it ends with an exception when run inside an x86 Android emulator.

We reported our findings to all three vendors and received varying responses. IBM assigned a CVE~\cite{bigdecimalcve} for our findings.  Oracle considered this a Security-in-Depth issue and acknowledged our contribution in their Critical Patch Update Advisory~\cite{securityindepthissue}. Google argued that it does not fall within the definition of a security vulnerability for the Android platform.

\thesystem automatically constructs valid instances of \texttt{BigDecimal} and \texttt{BigInteger} that substantially slow down methods in both classes. For example, simply incrementing \texttt{1e2147483647} by 1 takes over an hour to compute on Oracle's JDK even with Just-in-Time (JIT) Compilation enabled. \thesystem finds these vulnerabilities without any domain-specific knowledge about the Java Math library or the semantics of its classes; \thesystem derived all instances required to invoke methods by starting from the \texttt{BigDecimal} constructors given in the JRE.

The underlying issue in the JRE source code that introduces this vulnerability
stems from how it handles numbers expressed in scientific notation.  Every
number in scientific notation is expressed as a coefficient multiplied by ten
raised to the power of an exponent.  The performance of arithmetic over these
numbers in the JRE is sensitive to the difference between two numbers'
exponents.  This makes addition over two numbers with equal exponents, such as
\texttt{1e2147483647} and \texttt{2e2147483647}, return immediately, whereas
adding \texttt{1e2147483647} to \texttt{1e0}, can run for over an hour on
Oracle's JVM.

The root cause of this performance overhead lies in how the JDK transforms
numbers during its arithmetic operations~\cite{bigdecimalsrc}.  For example,
suppose a program uses the \texttt{BigDecimal} class to compute the sum \(x_1
\times 10^{y_1} + x_2 \times 10^{y_2}\) where \(y_1 \neq y_2\). Let \(x_{min}\)
be the coefficient that belongs to the smaller exponent and \(x_{max}\) the
coefficient that belongs to the larger exponent.  The \texttt{add} method first
computes the difference \(|y_1-y_2|\) and then defines an integer
\(x_{scaled}\), an instance of \texttt{BigInteger} which may represent an
integer of arbitrary size, and directly computes \(x_{scaled} = x_{max} \times
10^{|y_1-y_2|}\).  This allows the \texttt{add} method to complete the addition
by returning a \texttt{BigDecimal} represented with a coefficient given by
\(x_{scaled} + x_{min}\) and an exponent given by the smaller of \(y_1\) and
\(y_2\).  Unfortunately, it also opens up the possibility of a significant
performance penalty while computing \(x_{scaled}\) that an adversary could
exploit to achieve DoS when the difference between \(y_1\) and \(y_2\) is
large. In the PoC given above, \texttt{add} must compute \(x_{scaled} = 1
\times 10^{2147483647}\) before simply adding it to \(1\) with an exponent of
\(0\). Anecdotally, the \eyevm helped pinpoint this issue by tracing the
execution of the PoC. When tracing \(1 + 1 \times 10^{2147483647}\), the method
\texttt{bigMultiplyPowerTen} started computing \(1 \times 10^{2147483647}\) but
did not return before we manually stopped the PoC. This method appeared to be the
source of the performance penalty because it was absent when tracing
\(2 \times 10^{2147483647} + 1 \times 10^{2147483647}\) which completed
immediately.

After observing this result, we surveyed popular libraries that use
\texttt{BigDecimal} internally, and developed proof of concepts that exploit
this vulnerability as shown in Figure~\ref{fig:bigdecimalpoc}.  We found that
several general purpose programming languages hosted on the JVM are vulnerable
to this attack along with org.json, a popular JSON parsing library.

Developers face numerous security threats when they validate input values given
as strings. The vulnerabilities we discussed in this section are especially
problematic because malicious input is perfectly valid, albeit very large,
floating point numbers. If a program performs any arithmetic over a
\texttt{BigDecimal} object derived from user input, then it must take care to
prevent the user from providing arbitrary numbers in scientific notation.
Likewise, these results show that developers must be careful when converting
between these two classes, as interpreting certain floating point numbers as
integers could suddenly halt their application.  This complicates any input
validation that accepts numbers given as strings. Our results reveal that
failure to implement such validation correctly could allow remote adversaries
to slow victim programs to a halt.

After we disclosed this vulnerability to vendors, we observed that recent
releases of the JRE provide mitigations for it. For example, in JRE build
\texttt{1.8.0\_242}, the PoCs we present in this section
immediately end with an exception thrown by the \texttt{BigInteger} class.
Instead of naively computing \(x_{scaled}\), the new implementation first
checks to see if \(x_{scaled}\) will exceed a threshold and, if so, aborts the
inefficient computation with an exception before it can affect the availability
of the overall process. While this defends against the original AC vulnerability,
it also introduces a new opportunity for DoS by allowing an adversary to
trigger exceptions that programs may fail to properly handle.

\subsubsection{Case Study: DARPA STAC Challenges}
\label{subsec:darpa_eval}

The \stac program contains a set of challenge programs developed in Java that
test the ability of program analysis tools to detect AC vulnerabilities. In
this case study, we measure \thesystem's ability to automatically detect AC
vulnerabilities found in these challenges. We began by feeding the challenges
into \thesystem which produced a corpus of test cases that exploit AC
vulnerabilities contained in the challenges.

However, these test cases on their own do not answer the question of whether a
given challenge is vulnerable to an AC attack, because challenges are whole
programs that receive input from sources such as standard input or network
sockets, and \thesystem detects AC vulnerabilities at the method level.
Therefore, given a challenge that is vulnerable to an AC attack, we need a way
to determine whether one of its methods that \thesystem marks as vulnerable is
relevant to the intended vulnerability.

The STAC challenges provide ground truth for evaluating \thesystem in the form
of proof-of-concept exploits on challenges with intended vulnerabilities.
We define the following procedure to assess whether \thesystem can detect an AC
vulnerability automatically.
We start by executing each challenge that contains an intended vulnerability in
the \eyevm in tracing mode, and execute the challenge's exploit on the running
challenge. This produces a trace of every method invoked in the challenge
during a successful exploit. If \thesystem marks a method $M$ as vulnerable in
the output for challenge $C$, and $M$ appears in the trace for $C$, we count
the challenge as confirmed.

\begin{figure}[]
  \centering
    \includegraphics[angle=270,width=0.5\textwidth]{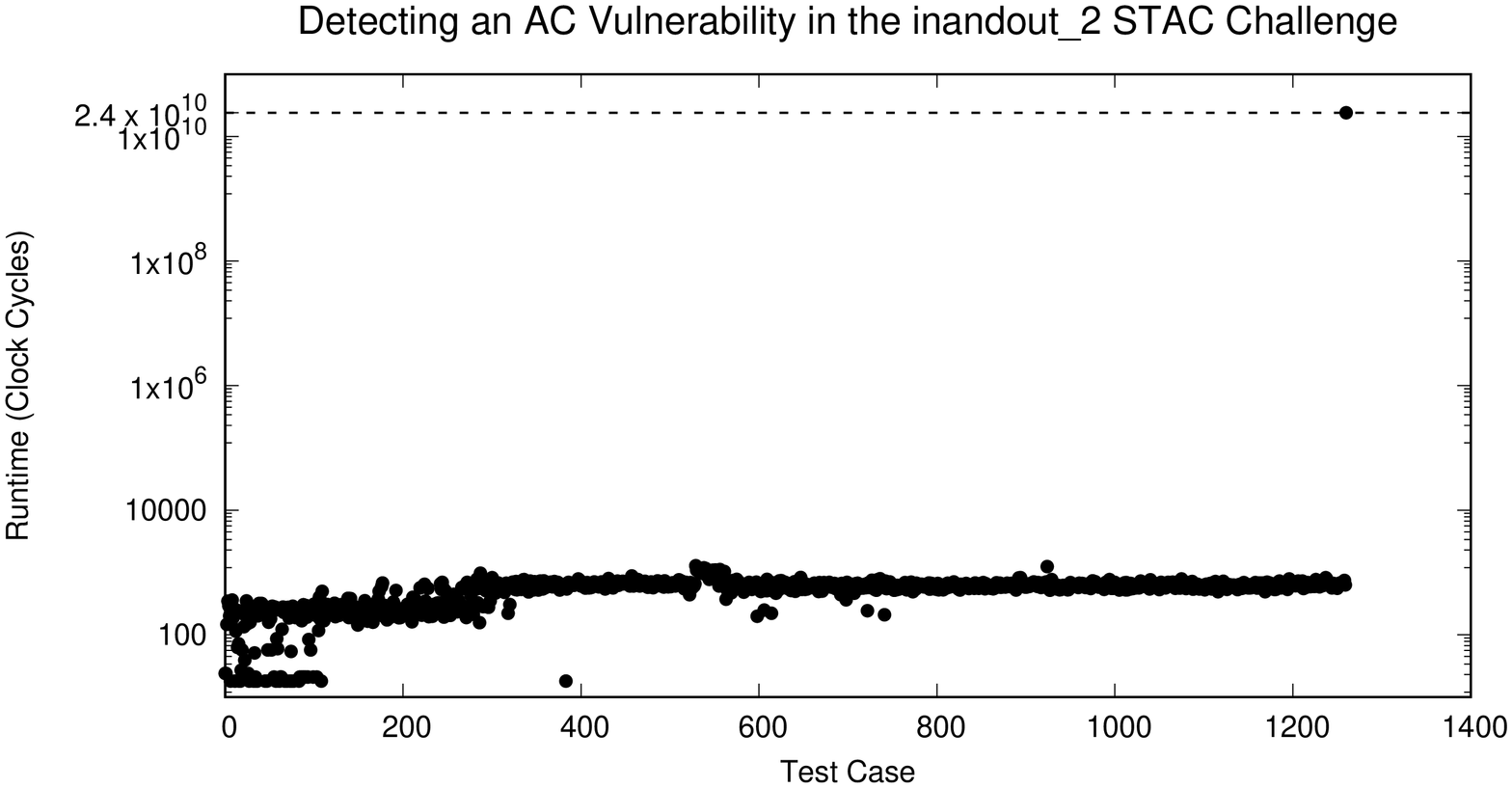}
\caption{Visualization of \thesystem detecting an AC vulnerability found inside the
inandout\_2 challenge in the DARPA STAC Program. The test case that triggers
the AC vulnerability in the challenge can be found in the upper-right-hand corner.}
    \label{fig:stac_challenge_ac}
\end{figure}

When we conducted this experiment on all the challenges contained in the STAC
program vulnerable to AC attacks, we found that \thesystem automatically marked
\solvedchallenges out of \totalchallenges challenges as vulnerable.
One challenge, \texttt{inandout\_2} provides a web service that allows users to
order pizzas online. By running \thesystem over this challenge, it identifies
multiple methods with AC vulnerabilities in its code.
Figure~\ref{fig:stac_challenge_ac} visualizes \thesystem detecting an AC
vulnerability in the \texttt{PizzaParameters.subsequentEnergyOf2(int)} method
found in the challenge. When we traced the execution of an exploit that achieved
DoS against the pizza service, we observed the vulnerable method
\texttt{subsequentEnergyOf2} identified by \thesystem in the exploit's trace.

\begin{figure}[]
    \centering
    \begin{lstlisting}[
      language=Java,
      numbers=left,
      stepnumber=1,
      xleftmargin=6mm,
      framexleftmargin=6mm
    ]
import com.sun.org.apache.bcel.internal.*;

Utility.replace("", "", "");

import java.io.File;
import sun.tools.jar.Manifest;

String xs[] = {"","","","","",""};
m = new Manifest();
files = new File(new File(""),"");
m.addFiles(files, xs);

import com.sun.imageio.plugins.common.*;

table = new LZWStringTable();
table.addCharString(0, 0);

import com.sun.org.apache.bcel.internal.*;

byte y[] = {0, 0, 0};
il = new InstructionList(y);
ifi = new InstructionFinder(il);
ifi.search("");

import sun.text.SupplementaryCharacterData;

int z[] = {0, 0, 0};
s = new SupplementaryCharacterData(z);
s.getValue(0);

import java.text.DecimalFormat;

x = new DecimalFormat("");
x.toLocalizedPattern();
\end{lstlisting}
    \caption{Proof of concept exploits for AC vulnerabilities that require only \ivi-based inputs to trigger.}
    \label{fig:emptyexploits}
\end{figure}

\subsubsection{Case Study: Slow Parsing in org.json}
\label{subsec:org_json}

Over the course of our evaluation \thesystem detected a previously unknown AC
vulnerability inside the popular org.json library. The vulnerable method,
\texttt{JSONML.toJSONObject(String)} converts an XML document represented as a
string into an equivalent JSONObject. This method is public, and instructions
for its use in programs can be found in tutorials online~\cite{jsontutorial}.
Given the popularity of the org.json library on Maven, application developers
may unknowingly expose themselves to DoS attacks by simply parsing XML strings
into JSON.

Our experimental results obtained by micro-fuzzing the \texttt{org.json}
library also demonstrated the utility of using \sri seed inputs over \ivi seed
inputs.  Over the course of our evaluation, test cases evolved from \ivi seed
inputs failed to successfully invoke the \texttt{toJSONObject} method after
4,654 test cases. Meanwhile, the 96th \sri-derived seed input successfully
triggered the vulnerability. The second stage of our pipeline successfully
validated this \sri test case represented as a 242 character string. After our
evaluation completed, we took the PoC program generated by \thesystem and
observed it running for \tojsonobjecttime on a Debian system with Intel Xeon
E5-2620 CPUs in an unmodified Java environment with JIT enabled. The \sri
strategy that produced this test case sampled primitive values uniformly at
random from the interval \([0, \alpha)\). Sampling from the normal distribution
\(\mathcal{N}(0, \nicefrac{\alpha}{3})\) detected a bug in the \texttt{JSONML}
package, but the test case did not pass the witness validation stage.

During our evaluation, \thesystem started with no prior knowledge about
org.json, JSON, or XML.  Nonetheless, after simply passing the org.json library
to \thesystem as input and micro-fuzzing its methods using \sri-derived seed
inputs, we were able to uncover a serious AC vulnerability that exposes
programs that depend on this library to potential DoS attacks.  We communicated
our findings to the owners of the \texttt{JSON-java} project~\cite{jsonjava}
who confirmed the behavior we observed as a bug.

The developers immediately found the root cause of the infinite loop after
debugging the method starting with the test case produced by \thesystem.
This test case opened an XML comment with the string \texttt{<!} which
prompted a loop inside the \texttt{toJSONObject} method to check for
special characters that represent the beginning (\texttt{<}) and end
(\texttt{>}) of a comment until it reached the end of the original comment. The
test case produced by \thesystem caused a method in this loop,
\texttt{nextMeta}, to always return the same character, and therefore prevented
the loop from advancing. After fixing this bug, the developers included our
test case in \texttt{org.json}'s test suite in order to prevent the issue from
occurring in future releases. The test case that triggered this infinite loop
is small (242 bytes) and demonstrates the potential micro-fuzzing has to
uncover serious AC vulnerabilities hiding in popular Java libraries. After
micro-fuzzing the \texttt{toJSONObject} method on the patched \texttt{org.json}
library, we discovered 6 test cases that triggered AC vulnerabilities, but
these were fixed in the latest release of \texttt{org.json}.

\section{Related Work}

\thesystem relates to previous work in four categories:
\begin{inparaenum}[\itshape (i)\upshape]
  \item AC Vulnerability Analysis,
  \item Test Case Generation,
  \item Fuzz Testing, and
  \item Resource Analysis.
\end{inparaenum}

\subsection{AC Vulnerability Analysis}

Prior work for detecting AC vulnerabilities in Java programs includes static
analysis on popular
libraries~\cite{wustholz_static_2017,kirrage13:regex_dos,livshits_finding_2005},
object-graph engineering on Java's serialization
facilities~\cite{dietrich_evil_2017}, and exploiting worst-case runtime of
algorithms found in commercial grade networking
equipment~\cite{czubak_algorithmic_2016}. On the Android platform,
Huang~\etal~\cite{huang_system_2015} use a combination of static and dynamic
analysis to detect AC vulnerabilities within Android's System Server. Further up
the application stack, Pellegrino~\etal~\cite{PellegrinoBWS15} identify common
implementation mistakes that make web services vulnerable to DoS attacks.
Finally, Holland~\etal~\cite{holland2016statically} proposes a 2 stage analysis
for finding AC vulnerabilities.

Prior work for detecting AC vulnerabilities is custom-tailored to specific
domains (e.g., serialization, regular-expression engines, Android Services, or
web applications) and therefore often requires human assistance. \thesystem
differs from these approaches in that it is generically applicable to any Java
program without human intervention, intuition, or insight.

\subsection{Test Case Generation}

Program analysis tools can generate test cases that exercise specific execution
paths in a program and demonstrate the presence of bugs. Several tools perform
symbolic execution within the Java Pathfinder~\cite{Havelund99}
platform~\cite{jayaraman_jfuzz:_2009, luckow_jdart:_2016, zhu_jfuzz:_2015} in
order to increase code coverage in Java test suites. Symbolic execution has
found serious security bugs when applied to whole
programs~\cite{molnar_dynamic_2009, cha_unleashing_2012} and in
under-constrained settings~\cite{ramos_under-constrained_2015} similar to
\thesystem. Toffola~\etal~\cite{ToffolaPG18} introduced PerfSyn which uses
combinatoric search to construct test programs that trigger performance
bottlenecks in Java methods.

\subsection{Fuzz Testing}

State of the art fuzzers~\cite{afl,libFuzzer} combine instrumentation on a
program under test to provide feedback to a genetic algorithm that mutates
inputs in order to trigger a crash. Active research topics include deciding
optimal fuzzing seeds~\cite{rebert_optimizing_2014} and techniques for
improving a fuzzer's code
coverage~\cite{cha_program-adaptive_2015,woo_scheduling_2013}. Prior work has
seeded fuzzing by replaying sequences of kernel API calls~\cite{HanC17},
commands from Android apps to smart IoT Devices~\cite{IOT18}, and input
provided by human assistants~\cite{shoshitaishvili2017rise}. Recent techniques
for improving code coverage during fuzz testing include introducing selective
symbolic execution~\cite{stephens_driller:_2016}, control- and data-flow
analysis on the program under test~\cite{rawat_vuzzer:_2017}, reducing
collisions in code coverage measurements~\cite{GanZQTLPC18}, and altering the
program under test~\cite{PengSP18}. Prior work applies existing fuzz testers to
discover AC vulnerabilities in whole programs~\cite{PetsiosZKJ17,LemieuxPSS18},
and in Java programs by combining fuzz testing with symbolic
execution~\cite{NollerKP18} or seeding black box fuzzing with information taken
from program traces~\cite{LuoNGP17}. In contrast, \thesystem micro-fuzzes
individual methods and uses a genetic algorithm on individual Java objects in
order to find inputs to these methods that demonstrate the presence of AC
vulnerabilities. This departs from prior approaches that restrict fuzzing
inputs to flat bitmaps.

\subsection{Resource Analysis}

Recent interest in AC and side-channel vulnerabilities increased the focus on
resource analysis research. In this area, Proteus~\cite{proteus} presented by
Xie~\etal and Awadhutkar~\etal~\cite{awdhutkar-apsec} study sensitive paths
through loops that might represent AC vulnerabilities. Meanwhile,
Kothary~\cite{human-machine,human-on-the-loop} investigates human-machine
interaction to improve program analysis for finding critical paths and side
channels. In Comb~\cite{Holland-comb}, Holland~\etal investigate how to improve
computation of all relevant program behaviors.

Other resource-oriented static analyses have also been proposed~\cite{Carbonneaux0RS17,Niari-2017}.
This line of work is based on statically inferred properties of programs and
their resource usage. In contrast, \thesystem provides quantitative measurements of program
behavior over concrete inputs in a dynamic, empirical fashion.

\section{Conclusion}
\label{sec:conclusion}

In this work, we present \thesystem, a fuzzer that detects Algorithmic
Complexity (AC) vulnerabilities in Java libraries through a novel approach
called \emph{micro-fuzzing}. \thesystem uses genetic optimization of test
artifact resource usage seeded by Java-specific Identity Value and Small
Recursive Instantiation (\ivi and \sri) techniques to search for inputs that
demonstrate AC vulnerabilities in methods under test.  We evaluate \thesystem
on the Java 8 Runtime Environment (JRE), challenge programs developed in the
DARPA Space and Time Analysis for Cyber-Security (STAC) program, and the 100
most popular libraries on Maven.  In conducting this evaluation, we discovered
previously unknown AC vulnerabilities in production software, including
\zerodaynum in the JRE, \mavenzerodaynum in \vulnmavenlibraries Maven
Libraries, as well as both known \emph{and} unintended vulnerabilities in STAC
evaluation artifacts. Our results demonstrate that the array of testing
techniques introduced by \thesystem are effective in finding AC vulnerabilities
in real-world software.

\section*{Acknowledgements}

This work was partially supported by the National Science Foundation (NSF)
under grant CNS-1703454 award, ONR grants N00014-19-1-2364 and
N00014-17-1-2011, and Secure Business Austria. The views and conclusions
contained herein are those of the authors and should not be interpreted as
representing the official policies or endorsements, either expressed or
implied, of any government agency.

{\footnotesize \bibliographystyle{acm}
\bibliography{hot-fuzz,misc}}

\begin{thebibliography}{10}

\bibitem{bigdecimalsrc}
{{Arithmetic in JDK BigDecimal}}.
\newblock
  \url{https://hg.openjdk.java.net/jdk8u/jdk8u-dev/jdk/file/d13abc740e42/src/share/classes/java/math/BigDecimal.java#l4464}.

\bibitem{bigdecimalcve}
{{CVE}}-2018-1517.
\newblock \url{http://cve.mitre.org/cgi-bin/cvename.cgi?name=CVE-2018-1517}.

\bibitem{linuxcve}
{{CVE}}-2018-5390.
\newblock
  \url{https://nvd.nist.gov/vuln/detail/CVE-2018-5390#vulnCurrentDescriptionTitle}.

\bibitem{golangcve}
{{CVE}}-2019-6486.
\newblock \url{http://cve.mitre.org/cgi-bin/cvename.cgi?name=CVE-2019-6486}.

\bibitem{jsonjava}
{{JSON-Java Project}}.
\newblock \url{https://stleary.github.io/JSON-java/}.

\bibitem{jsontutorial}
{JSONML Tutorials Point}.
\newblock \url{https://www.tutorialspoint.com/org_json/org_json_jsonml.htm}.

\bibitem{libFuzzer}
{libFuzzer – A library for coverage-guided fuzz testing}.
\newblock \url{https://llvm.org/docs/LibFuzzer.html}.

\bibitem{maven}
{Maven Repository}.
\newblock \url{https://mvnrepository.com/}.

\bibitem{securityindepthissue}
{Oracle Critical Patch Update Advisory - January 2019}.
\newblock
  \url{https://www.oracle.com/technetwork/security-advisory/cpujan2019-5072801.html}.

\bibitem{Java8Spec}
{The Java Virtual Machine Specification}.
\newblock \url{https://docs.oracle.com/javase/specs/jvms/se8/html/index.html}.

\bibitem{jvmti}
{The JVM Tool Interface (JVM TI): How VM Agents Work}.
\newblock
  \url{https://www.oracle.com/technetwork/articles/javase/index-140680.html}.

\bibitem{redqueen19}
{\sc Aschermann, C., Schumilo, S., Blazytko, T., Gawlik, R., and Holz, T.}
\newblock Redqueen: Fuzzing with input-to-state correspondence.
\newblock In {\em NDSS\/} (2019).

\bibitem{awdhutkar-apsec}
{\sc Awadhutkar, P., Santhanam, G.~R., Holland, B., and Kothari, S.}
\newblock Intelligence amplifying loop characterizations for detecting
  algorithmic complexity vulnerabilities.
\newblock In {\em 2017 24th Asia-Pacific Software Engineering Conference
  (APSEC)\/} (Dec. 2017), vol.~00, pp.~249--258.

\bibitem{netflix}
{\sc Behrens, S., and Payne, B.}
\newblock Starting the avalanche: Application ddos in microservice
  architectures.

\bibitem{Carbonneaux0RS17}
{\sc Carbonneaux, Q., Hoffmann, J., Reps, T.~W., and Shao, Z.}
\newblock Automated resource analysis with coq proof objects.
\newblock In {\em Computer Aided Verification - 29th International Conference,
  {CAV} 2017, Heidelberg, Germany, July 24-28, 2017, Proceedings, Part {II}\/}
  (2017), pp.~64--85.

\bibitem{cha_unleashing_2012}
{\sc Cha, S.~K., Avgerinos, T., Rebert, A., and Brumley, D.}
\newblock Unleashing {Mayhem} on {Binary} {Code}.
\newblock In {\em {IEEE} {Symposium} on {Security} and {Privacy}, {SP} 2012,
  21-23 {May} 2012, {San} {Francisco}, {California}, {USA}\/} (2012),
  pp.~380--394.

\bibitem{cha_program-adaptive_2015}
{\sc Cha, S.~K., Woo, M., and Brumley, D.}
\newblock Program-{Adaptive} {Mutational} {Fuzzing}.
\newblock In {\em 2015 {IEEE} {Symposium} on {Security} and {Privacy}, {SP}
  2015, {San} {Jose}, {CA}, {USA}, {May} 17-21, 2015\/} (2015), pp.~725--741.

\bibitem{IOT18}
{\sc Chen, J., Diao, W., Zhao, Q., Zuo, C., Lin, Z., Wang, X., Lau, W.~C., Sun,
  M., Yang, R., and Zhang, K.}
\newblock Iotfuzzer: Discovering memory corruptions in iot through app-based
  fuzzing.
\newblock In {\em 25th Annual Network and Distributed System Security
  Symposium, {NDSS} 2018, San Diego, California, USA, February 18-21, 2018\/}
  (2018).

\bibitem{crosby_denial_2003}
{\sc Crosby, S.~A., and Wallach, D.~S.}
\newblock Denial of {Service} via {Algorithmic} {Complexity} {Attacks}.
\newblock In {\em Proceedings of the 12th {USENIX} {Security} {Symposium},
  {Washington}, {D}.{C}., {USA}, {August} 4-8, 2003\/} (2003).

\bibitem{czubak_algorithmic_2016}
{\sc Czubak, A., and Szymanek, M.}
\newblock Algorithmic {Complexity} {Vulnerability} {Analysis} of a {Stateful}
  {Firewall}.
\newblock In {\em Information {Systems} {Architecture} and {Technology}:
  {Proceedings} of 37th {International} {Conference} on {Information} {Systems}
  {Architecture} and {Technology} - {ISAT} 2016 - {Part} {II}\/} (2016),
  pp.~77--97.

\bibitem{dietrich_evil_2017}
{\sc Dietrich, J., Jezek, K., Rasheed, S., Tahir, A., and Potanin, A.}
\newblock Evil {Pickles}: {DoS} {Attacks} {Based} on {Object}-{Graph}
  {Engineering}.
\newblock In {\em 31st {European} {Conference} on {Object}-{Oriented}
  {Programming}, {ECOOP} 2017, {June} 19-23, 2017, {Barcelona}, {Spain}\/}
  (2017), pp.~10:1--10:32.

\bibitem{EibenS15}
{\sc Eiben, A.~E., and Smith, J.~E.}
\newblock {\em Introduction to Evolutionary Computing}.
\newblock Natural Computing Series. Springer, 2015.

\bibitem{GanZQTLPC18}
{\sc Gan, S., Zhang, C., Qin, X., Tu, X., Li, K., Pei, Z., and Chen, Z.}
\newblock Collafl: Path sensitive fuzzing.
\newblock In {\em 2018 {IEEE} Symposium on Security and Privacy, {SP} 2018,
  Proceedings, 21-23 May 2018, San Francisco, California, {USA}\/} (2018),
  pp.~679--696.

\bibitem{godefroid2014}
{\sc Godefroid, P.}
\newblock Micro execution.
\newblock In {\em Proceedings of the 36th International Conference on Software
  Engineering\/} (2014), ACM, pp.~539--549.

\bibitem{HanC17}
{\sc Han, H., and Cha, S.~K.}
\newblock {IMF:} inferred model-based fuzzer.
\newblock In {\em Proceedings of the 2017 {ACM} {SIGSAC} Conference on Computer
  and Communications Security, {CCS} 2017, Dallas, TX, USA, October 30 -
  November 03, 2017\/} (2017), pp.~2345--2358.

\bibitem{Havelund99}
{\sc Havelund, K.}
\newblock Java pathfinder, {A} translator from java to promela.
\newblock In {\em Theoretical and Practical Aspects of {SPIN} Model Checking,
  5th and 6th International {SPIN} Workshops, Trento, Italy, July 5, 1999,
  Toulouse, France, September 21 and 24 1999, Proceedings\/} (1999), p.~152.

\bibitem{Holland-comb}
{\sc Holland, B., Awadhutkar, P., Kothari, S., Tamrawi, A., and Mathews, J.}
\newblock Comb: Computing relevant program behaviors.
\newblock In {\em Proceedings of the 40th International Conference on Software
  Engineering: Companion Proceeedings\/} (New York, NY, USA, 2018), ICSE '18,
  ACM, pp.~109--112.

\bibitem{holland2016statically}
{\sc Holland, B., Santhanam, G.~R., Awadhutkar, P., and Kothari, S.}
\newblock Statically-informed dynamic analysis tools to detect algorithmic
  complexity vulnerabilities.
\newblock In {\em Source Code Analysis and Manipulation (SCAM), 2016 IEEE 16th
  International Working Conference on\/} (2016), IEEE, pp.~79--84.

\bibitem{huang_system_2015}
{\sc Huang, H., Zhu, S., Chen, K., and Liu, P.}
\newblock From {System} {Services} {Freezing} to {System} {Server} {Shutdown}
  in {Android}: {All} {You} {Need} {Is} a {Loop} in an {App}.
\newblock In {\em Proceedings of the 22nd {ACM} {SIGSAC} {Conference} on
  {Computer} and {Communications} {Security}, {Denver}, {CO}, {USA}, {October}
  12-6, 2015\/} (2015), pp.~1236--1247.

\bibitem{jayaraman_jfuzz:_2009}
{\sc Jayaraman, K., Harvison, D., Ganesh, V., and Kiezun, A.}
\newblock {jFuzz}: {A} {Concolic} {Whitebox} {Fuzzer} for {Java}.
\newblock In {\em First {NASA} {Formal} {Methods} {Symposium} - {NFM} 2009,
  {Moffett} {Field}, {California}, {USA}, {April} 6-8, 2009.\/} (2009),
  pp.~121--125.

\bibitem{isstac}
{\sc Kersten, R.}
\newblock {Kelinci}.
\newblock \url{https://github.com/isstac/kelinci}.

\bibitem{kirrage13:regex_dos}
{\sc Kirrage, J., Rathnayake, A., and Thielecke, H.}
\newblock Static analysis for regular expression denial-of-service attacks.
\newblock In {\em Proceedings of the International Conference on Network and
  System Security ({NSS})\/} (Madrid, Spain, June 2013).

\bibitem{KleesRCW018}
{\sc Klees, G., Ruef, A., Cooper, B., Wei, S., and Hicks, M.}
\newblock Evaluating fuzz testing.
\newblock In {\em Proceedings of the 2018 {ACM} {SIGSAC} Conference on Computer
  and Communications Security, {CCS} 2018, Toronto, ON, Canada, October 15-19,
  2018\/} (2018), pp.~2123--2138.

\bibitem{human-machine}
{\sc {Kothari}, S., {Tamrawi}, A., and {Mathews}, J.}
\newblock Human-machine resolution of invisible control flow?
\newblock In {\em 2016 IEEE 24th International Conference on Program
  Comprehension (ICPC)\/} (May 2016), pp.~1--4.

\bibitem{kuleshov2007using}
{\sc Kuleshov, E.}
\newblock Using the asm framework to implement common java bytecode
  transformation patterns.
\newblock {\em Aspect-Oriented Software Development\/} (2007).

\bibitem{laf}
{\sc laf intel}.
\newblock {Circumventing Fuzzing Roadblocks with Compiler Transformations}.
\newblock \url{https://lafintel.wordpress.com/}.

\bibitem{LemieuxPSS18}
{\sc Lemieux, C., Padhye, R., Sen, K., and Song, D.}
\newblock Perffuzz: automatically generating pathological inputs.
\newblock In {\em Proceedings of the 27th {ACM} {SIGSOFT} International
  Symposium on Software Testing and Analysis, {ISSTA} 2018, Amsterdam, The
  Netherlands, July 16-21, 2018\/} (2018), pp.~254--265.

\bibitem{livshits_finding_2005}
{\sc Livshits, V.~B., and Lam, M.~S.}
\newblock Finding {Security} {Vulnerabilities} in {Java} {Applications} with
  {Static} {Analysis}.
\newblock In {\em Proceedings of the 14th {USENIX} {Security} {Symposium},
  {Baltimore}, {MD}, {USA}, {July} 31 - {August} 5, 2005\/} (2005).

\bibitem{luckow_jdart:_2016}
{\sc Luckow, K.~S., Dimjasevic, M., Giannakopoulou, D., Howar, F., Isberner,
  M., Kahsai, T., Rakamaric, Z., and Raman, V.}
\newblock {JDart}: {A} {Dynamic} {Symbolic} {Analysis} {Framework}.
\newblock In {\em Tools and {Algorithms} for the {Construction} and {Analysis}
  of {Systems} - 22nd {International} {Conference}, {TACAS} 2016, {Held} as
  {Part} of the {European} {Joint} {Conferences} on {Theory} and {Practice} of
  {Software}, {ETAPS} 2016, {Eindhoven}, {The} {Netherlands}, {April} 2-8,
  2016, {Proceedings}\/} (2016), pp.~442--459.

\bibitem{LuoNGP17}
{\sc Luo, Q., Nair, A., Grechanik, M., and Poshyvanyk, D.}
\newblock {FOREPOST:} finding performance problems automatically with
  feedback-directed learning software testing.
\newblock {\em Empirical Software Engineering 22}, 1 (2017), 6--56.

\bibitem{msfuzz}
{\sc Microsoft}.
\newblock Microsoft security risk detection.
\newblock \url{https://www.microsoft.com/en-us/security-risk-detection/}.

\bibitem{molnar_dynamic_2009}
{\sc Molnar, D., Li, X.~C., and Wagner, D.~A.}
\newblock Dynamic {Test} {Generation} to {Find} {Integer} {Bugs} in x86
  {Binary} {Linux} {Programs}.
\newblock In {\em 18th {USENIX} {Security} {Symposium}, {Montreal}, {Canada},
  {August} 10-14, 2009, {Proceedings}\/} (2009), pp.~67--82.

\bibitem{NollerKP18}
{\sc Noller, Y., Kersten, R., and Pasareanu, C.~S.}
\newblock Badger: complexity analysis with fuzzing and symbolic execution.
\newblock In {\em Proceedings of the 27th {ACM} {SIGSOFT} International
  Symposium on Software Testing and Analysis, {ISSTA} 2018, Amsterdam, The
  Netherlands, July 16-21, 2018\/} (2018), pp.~322--332.

\bibitem{munch18}
{\sc Ognawala, S., Hutzelmann, T., Psallida, E., and Pretschner, A.}
\newblock Improving function coverage with munch: a hybrid fuzzing and directed
  symbolic execution approach.
\newblock In {\em Proceedings of the 33rd Annual ACM Symposium on Applied
  Computing\/} (2018), ACM, pp.~1475--1482.

\bibitem{PellegrinoBWS15}
{\sc Pellegrino, G., Balzarotti, D., Winter, S., and Suri, N.}
\newblock In the compression hornet's nest: {A} security study of data
  compression in network services.
\newblock In {\em 24th {USENIX} Security Symposium, {USENIX} Security 15,
  Washington, D.C., USA, August 12-14, 2015.\/} (2015), pp.~801--816.

\bibitem{PengSP18}
{\sc Peng, H., Shoshitaishvili, Y., and Payer, M.}
\newblock T-fuzz: Fuzzing by program transformation.
\newblock In {\em 2018 {IEEE} Symposium on Security and Privacy, {SP} 2018,
  Proceedings, 21-23 May 2018, San Francisco, California, {USA}\/} (2018),
  pp.~697--710.

\bibitem{PetsiosZKJ17}
{\sc Petsios, T., Zhao, J., Keromytis, A.~D., and Jana, S.}
\newblock Slowfuzz: Automated domain-independent detection of algorithmic
  complexity vulnerabilities.
\newblock In {\em Proceedings of the 2017 {ACM} {SIGSAC} Conference on Computer
  and Communications Security, {CCS} 2017, Dallas, TX, USA, October 30 -
  November 03, 2017\/} (2017), pp.~2155--2168.

\bibitem{ramos_under-constrained_2015}
{\sc Ramos, D.~A., and Engler, D.~R.}
\newblock Under-{Constrained} {Symbolic} {Execution}: {Correctness} {Checking}
  for {Real} {Code}.
\newblock In {\em 24th {USENIX} {Security} {Symposium}, {USENIX} {Security} 15,
  {Washington}, {D}.{C}., {USA}, {August} 12-14, 2015.\/} (2015), pp.~49--64.

\bibitem{guidovranken}
{\sc Ranken, G.}
\newblock {libFuzzer Java}.
\newblock \url{https://github.com/guidovranken/libfuzzer-java}.

\bibitem{rawat_vuzzer:_2017}
{\sc Rawat, S., Jain, V., Kumar, A., Cojocar, L., Giuffrida, C., and Bos, H.}
\newblock Vuzzer: {Application}-aware evolutionary fuzzing.
\newblock In {\em Proceedings of the {Network} and {Distributed} {System}
  {Security} {Symposium} ({NDSS})\/} (2017).

\bibitem{rebert_optimizing_2014}
{\sc Rebert, A., Cha, S.~K., Avgerinos, T., Foote, J., Warren, D., Grieco, G.,
  and Brumley, D.}
\newblock Optimizing {Seed} {Selection} for {Fuzzing}.
\newblock In {\em Proceedings of the 23rd {USENIX} {Security} {Symposium},
  {San} {Diego}, {CA}, {USA}, {August} 20-22, 2014.\/} (2014), pp.~861--875.

\bibitem{RussellB01}
{\sc Russell, K.~B., and Bak, L.}
\newblock The hotspot serviceability agent: An out-of-process high-level
  debugger for a java virtual machine.
\newblock In {\em Proceedings of the 1st Java Virtual Machine Research and
  Technology Symposium, April 23-24, 2001, Monterey, CA, {USA}\/} (2001),
  pp.~117--126.

\bibitem{human-on-the-loop}
{\sc Santhanam, G.~R., Holland, B., Kothari, S., and Ranade, N.}
\newblock Human-on-the-loop automation for detecting software side-channel
  vulnerabilities.
\newblock In {\em Information Systems Security\/} (Cham, 2017), R.~K.
  Shyamasundar, V.~Singh, and J.~Vaidya, Eds., Springer International
  Publishing, pp.~209--230.

\bibitem{asan12}
{\sc Serebryany, K., Bruening, D., Potapenko, A., and Vyukov, D.}
\newblock Addresssanitizer: A fast address sanity checker.
\newblock In {\em Presented as part of the 2012 USENIX Annual Technical
  Conference (USENIX ATC 12)\/} (2012), pp.~309--318.

\bibitem{shoshitaishvili2017rise}
{\sc Shoshitaishvili, Y., Weissbacher, M., Dresel, L., Salls, C., Wang, R.,
  Kruegel, C., and Vigna, G.}
\newblock Rise of the hacrs: Augmenting autonomous cyber reasoning systems with
  human assistance.
\newblock In {\em Proceedings of the 2017 ACM Conference on Computer and
  Communications Security\/} (2017), ACM.

\bibitem{ossfuzz}
{\sc Source, G.~O.}
\newblock Oss-fuzz.
\newblock \url{https://github.com/google/oss-fuzz}.

\bibitem{StaicuP18}
{\sc Staicu, C., and Pradel, M.}
\newblock Freezing the web: {A} study of redos vulnerabilities in
  javascript-based web servers.
\newblock In {\em 27th {USENIX} Security Symposium, {USENIX} Security 2018,
  Baltimore, MD, USA, August 15-17, 2018.\/} (2018), pp.~361--376.

\bibitem{stephens_driller:_2016}
{\sc Stephens, N., Grosen, J., Salls, C., Dutcher, A., Wang, R., Corbetta, J.,
  Shoshitaishvili, Y., Kruegel, C., and Vigna, G.}
\newblock Driller: {Augmenting} {Fuzzing} {Through} {Selective} {Symbolic}
  {Execution}.
\newblock Internet Society.

\bibitem{Niari-2017}
{\sc Tizpaz-Niari, S., {\v{C}}ern{\'y}, P., Chang, B.-Y.~E., Sankaranarayanan,
  S., and Trivedi, A.}
\newblock Discriminating traces with time.
\newblock In {\em Tools and Algorithms for the Construction and Analysis of
  Systems\/} (Berlin, Heidelberg, 2017), A.~Legay and T.~Margaria, Eds.,
  Springer Berlin Heidelberg, pp.~21--37.

\bibitem{ToffolaPG18}
{\sc Toffola, L.~D., Pradel, M., and Gross, T.~R.}
\newblock Synthesizing programs that expose performance bottlenecks.
\newblock In {\em Proceedings of the 2018 International Symposium on Code
  Generation and Optimization, {CGO} 2018, V{\"{o}}sendorf / Vienna, Austria,
  February 24-28, 2018\/} (2018), pp.~314--326.

\bibitem{woo_scheduling_2013}
{\sc Woo, M., Cha, S.~K., Gottlieb, S., and Brumley, D.}
\newblock {Scheduling} {Black}-{Box} {Mutational} {Fuzzing}.
\newblock In {\em 2013 {ACM} {SIGSAC} {Conference} on {Computer} and
  {Communications} {Security}, {CCS}'13, {Berlin}, {Germany}, {November} 4-8,
  2013\/} (2013), pp.~511--522.

\bibitem{wustholz_static_2017}
{\sc Wüstholz, V., Olivo, O., Heule, M. J.~H., and Dillig, I.}
\newblock Static {Detection} of {DoS} {Vulnerabilities} in {Programs} that
  {Use} {Regular} {Expressions}.
\newblock In {\em Tools and {Algorithms} for the {Construction} and {Analysis}
  of {Systems} - 23rd {International} {Conference}, {TACAS} 2017, {Held} as
  {Part} of the {European} {Joint} {Conferences} on {Theory} and {Practice} of
  {Software}, {ETAPS} 2017, {Uppsala}, {Sweden}, {April} 22-29, 2017,
  {Proceedings}, {Part} {II}\/} (2017), pp.~3--20.

\bibitem{proteus}
{\sc Xie, X., Chen, B., Liu, Y., Le, W., and Li, X.}
\newblock Proteus: Computing disjunctive loop summary via path dependency
  analysis.
\newblock In {\em Proceedings of the 2016 24th ACM SIGSOFT International
  Symposium on Foundations of Software Engineering\/} (New York, NY, USA,
  2016), FSE 2016, ACM, pp.~61--72.

\bibitem{afl}
{\sc Zalewski, M.}
\newblock {American Fuzzy Lop}.
\newblock \url{http://lcamtuf.coredump.cx/afl/}.

\bibitem{zhu_jfuzz:_2015}
{\sc Zhu, H.}
\newblock {JFuzz}: {A} {Tool} for {Automated} {Java} {Unit} {Testing} {Based}
  on {Data} {Mutation} and {Metamorphic} {Testing} {Methods}.
\newblock In {\em 2015 {Second} {International} {Conference} on {Trustworthy}
  {Systems} and {Their} {Applications}, {TSA} 2015, {Hualien}, {Taiwan}, {July}
  8-9, 2015\/} (2015), pp.~8--15.

\end{thebibliography}

\end{document}